\documentclass{article}

\usepackage{authblk} %Prepare Author and Affiliation Blocks
\usepackage[bookmarksnumbered=true,
	   	 colorlinks=true,
            linkcolor=blue,
            citecolor=red,
            urlcolor=blue
            ]{hyperref}
\usepackage[left=2.5cm, right=2.5cm, top=2.5cm, bottom=2.2cm]{geometry}
\usepackage{natbib} % reimplementation of the LATEX \cite command,
\usepackage{bm}
\usepackage{amsfonts,amssymb} % provide some mathematical fonts by the AMS
\usepackage{amsmath} % provide several extensions to typeset mathematics
\usepackage{commath} % provide some commands which are supposed to produce better formatting and/or provide a more convenient way of input.
\usepackage{xcolor}
\usepackage{tikz}
\usepackage{overpic}
\usepackage{epstopdf}
\usepackage[titletoc,title]{appendix}% Extra control of appendices, enable the appendices environment
\numberwithin{equation}{section} % prepend the section number to all equation numbers
\usepackage[noabbrev]{cleveref} % enhances LATEX's cross-referencing features, allowing the format of references to be determined automatically according to the type of reference.
\crefname{appsec}{appendix}{appendices} % define appsec as appendix for singular and plural
\begin{document}

\title{Drop spreading and drifting on a spatially heterogeneous film: \\ capturing variability with asymptotics and emulation}
\author{Feng Xu$^1$\footnote{f.xu.2@bham.ac.uk}}
\author{Sam Coveney$^2$}
\author{Oliver E. Jensen$^3$\footnote{oliver.jensen@manchester.ac.uk}}
\affil{$^1$ School of Mathematics, University of Birmingham, Edgbaston B15 2TT}
\affil{$^2$ Department of Physics \& Astronomy, University of Sheffield, \\ Hounsfield Road, Sheffield S3 7RH, UK}
\affil{$^3$ School of Mathematics, University of Manchester, \\ Oxford Road, Manchester M13 9PL, UK}
\date{\today}
\maketitle

\begin{abstract}
A liquid drop spreading over a thin heterogeneous precursor film (such as an inhaled droplet on the mucus-lined wall of a lung airway) will experience perturbations in shape and location as its advancing contact line encounters regions of low or high film viscosity.  Prior work on spatially one-dimensional spreading over a precursor film having a random viscosity field [Xu \& Jensen 2016, Proc. Roy. Soc. A 472, 20160270] has demonstrated how viscosity fluctuations are swept into a narrow region behind the {\color{black}drop's effective} contact line, where they can impact drop dynamics.  Here we investigate two-dimensional drops, seeking to understand the relationship between the statistical properties of the precursor film and those of the spreading drop.  Assuming the precursor film is much thinner than the drop and viscosity fluctuations are weak, we use asymptotic methods to derive explicit predictions for the mean and variance of drop area and the drop's lateral drift.  For larger film variability, we use Gaussian process emulation to estimate the variance of outcomes from a restricted set of simulations.  Stochastic drift of the droplet is predicted to be greatest when the initial drop diameter is comparable to the correlation length of viscosity fluctuations.
\end{abstract}

\section{Introduction}

Natural materials commonly have spatially non-uniform properties.  Such features can often be conveniently characterised as stochastic fields, whereby statistical parameters are used to represent inherent randomness or to summarise limited knowledge of the structure of an individual specimen.  Transport processes influenced by such heterogeneous fields can exhibit features that may be overlooked in `clean' engineered experiments, and traditional modelling approaches must be modified to predict the distribution of potential outcomes arising from material variability.  Here we investigate a system in which spatial disorder influences, and is influenced by, a canonical thin-film viscous flow.  

We consider a model problem motivated by a process arising in the human lung, whereby an inhaled aerosol droplet interacts with the mucus film lining an airway. The droplet may contain a virus or drug that is delivered to the film and the underlying tissue.  Mucus is a complex material containing mucin proteins \citep{thornton2008,lai2009,widdicombe2015} {\color{black}that endow the film with spatially heterogeneous properties.  Mucins} are released from goblet cells {\color{black}and submucosal glands} interspersed across the airway wall, modifying the {\color{black} mucus rheology in a manner leading to a high degree of variability within and between individuals \citep{sims1997, kirkham2002, levy2014}; microrheological measurements also reveal variations across time- and length-scales \citep{lai2009}.  \textit{En face} imaging of excised airway epithelium reveals the presence of discrete sheets of the mucin MUC5AC and threads of the mucin MUC5B that can extend over hundreds of microns \citep{ostedgaard2017}, in addition to localised mucin patches appearing on a roughly 10$\mu$m lengthscale associated with release sites; the patches appear as ``clouds" separated by ``plumes" of proteoglycan when imaged transverse to the plane of the mucus layer \citep{kesimer2013}.  While numerous additional factors will influence the spreading of a drop over a mucus film, not least the film's complex non-Newtonian rheology, these observations of mucin heterogeneity motivate a considerably simpler but nevertheless fundamental question, namely how in-plane} spatial {\color{black}gradients} in the viscosity of an endogenous {\color{black}liquid} layer lead to variations in the {\color{black}area of a drop spreading over it}, and how inhomogeneity leads to lateral drift of the drop as it spreads.  The spatial disorder of the film demands that drop spreading is defined by quantities such as the mean and variance (over multiple realisations) of the drop area and the location of its centre of mass. 

In our approach to this {\color{black}drop-spreading} problem, the disordered viscosity field of the mucus layer on which the droplet lands provides a stochastic initial condition to an otherwise deterministic model.  This model tracks the resulting evolution of the fluid layer and that of a solute (a mucin proxy) that determines the evolving viscosity field; {\color{black}like \cite{chatelin2016}, who modelled mucociliary clearance, we treat the film as Newtonian but heterogeneous.}  The model's coupled nonlinear evolution equations can be solved repeatedly, for multiple realisations of the initial conditions, to determine the distribution of outcomes.  However direct Monte Carlo (MC) simulation can become prohibitively expensive if variance is to be calculated accurately over a wide range of input parameters. A key goal of this study is to develop and illustrate more economical approaches to uncertainty quantification, {\color{black}which we implement using Gaussian process emulation and model reduction}.  

We have previously addressed this problem in one spatial dimension \citep{2016-Xu-vol472}, showing how an asymptotic {\color{black}approximation} of the PDE system yields a low-order surrogate model that is amenable to repeated simulation and further analysis, allowing explicit evaluation of output variances.  We extend this approach from one to two spatial dimensions below, identifying a novel free-boundary problem along the way.  While the asymptotic approximation is helpful in revealing dominant physical processes, slow convergence of a logarithmic series restricts its accuracy.  Accordingly, we also deploy a complementary numerical approach whereby we emulate the computational model as a Gaussian process (GP), using a restricted set of training data (a set of simulations across a 2D input parameter space, evaluated using a structured design algorithm) to estimate variances efficiently for a wider range of parameter values.

We focus attention on the situation in which the initial drop height is substantially greater than the thickness of the undisturbed precursor film.  {\color{black}The film regularizes the singularity associated with a moving contact line, obviating any requirement to include the effects of slip; the drop is assumed to be fully miscible with the film and for simplicity disjoining pressure effects are neglected.}  Spreading is then regulated primarily by flows confined to narrow regions near the {\color{black} junction between the bulk of the drop and the precursor film, which we will refer to as the drop's effective} contact line.  The solute (or viscosity) field is stretched by the spreading flow beneath the drop, but strongly compressed immediately behind the {\color{black}effective} contact line.  The short lengthscales in pressure and solute fields near the contact line make simulations computationally expensive, particularly in two spatial dimensions, although they allow an asymptotic approximation to be developed.  In particular, we exploit the fact that over early to moderate times, the viscosity field sampled by the initial contact line retains a long-lived influence on spreading rates \citep{2016-Xu-vol472}.  We also consider a further limit in which the disorder in the initial viscosity field is weak, so that fluctuations satisfy a linear transport problem.  This enables us to derive explicit predictions {\color{black}for quantities such as the direction of lateral drift of the drop in a particular realisation, the mean-square displacement of the drop over multiple realisations, and} the variance of drop dimensions as a function of the correlation length of the initial solute field.  

The present study complements studies of thin films spreading in one spatial dimension on rough or textured surfaces \citep{cox1983, miksis1994, savva2010}, but differs from these through the fact that the spatial disorder in the problem evolves with the flow, as well as through consideration of features specific to two spatial dimensions.  We deliberately ignore many of the complexities of lung airways, such as non-Newtonian rheology, surfactant, stresses due to airflow, substrate curvature, {\color{black}the presence of cilia and periciliary fluid beneath the mucus layer,} etc. \citep{levy2014}, instead addressing a simplified model problem that involves a dominant balance between uniform surface tension and viscous effects.  The model is formulated in Sec.~\ref{sec:2} and simulations are presented in Sec.~\ref{sec:sim}, where a GP emulator is used to characterise the variance of outputs across parameter space.  The model's asymptotic reduction is derived in Sec.~\ref{sec:3}, and variance predictions are evaluated in comparison to emulation results.  {\color{black}To avoid duplication, figures presented in Sec.~\ref{sec:sim} contain results from both Sec.~\ref{sec:sim} and \ref{sec:3}.}

\section{The model problem and methods}
\label{sec:2}

We consider a thin film of Newtonian liquid with spatially heterogeneous viscosity on a horizontal plate. We assume the liquid's viscosity is linearly proportional to the concentration of a chemical species in the liquid which is passively transported.  The solute is treated as a large molecular-weight polymer with a concentration that quickly comes into equilibrium across the thin liquid layer, but diffuses relatively slowly along it, so that heterogeneities in its concentration persist over long time intervals. The lower surface of the liquid satisfies the no-slip and no-penetration conditions and the upper surface is subject to uniform surface tension.  Using lubrication theory and cross-film averaging, as in \cite{2016-Xu-vol472}, the evolution of the liquid film satisfies the dimensionless equations
\begin{align}
\label{2d_vector_form}
H_t + \nabla \cdot (H \bm{u}) = 0, \quad M_t + \bm{u} \cdot \nabla M = \frac{1}{Pe}\frac{\nabla \cdot (H \nabla M)}{H}, \quad \bm{u} \equiv \frac{H^2}{3M} \nabla \nabla^2 H.
\end{align}
Here $H(x,y,t)$ is the film thickness (scaled on $h_0$), $M(x, y, t)$ the cross-sectionally averaged viscosity field (scaled on $\mu_0$), $t$ time (scaled on $\mu_0 l_0^4/h_0^3 \gamma$), $x$ and $y$ the planar Cartesian coordinates (scaled on $l_0$) and $\nabla$ the gradient operator. Later, we also use planar polar coordinates $(r, \theta)$. $\mu_0$ is the mean viscosity, $l_0$ is the initial drop {\color{black}width}, $h_0$ $(\ll l_0)$ the initial drop height and $\gamma$ the uniform surface tension.  The PDEs in (\ref{2d_vector_form}) represent conservation of mass for the fluid and transport of the solute $M$; $\bm{u}$ represents an in-plane velocity, proportional to gradients of the capillary pressure $-\nabla^2 H$.   Letting $D$ {\color{black}denote} the molecular diffusivity, we choose the Pe\'clet number $Pe=h_0^3\gamma/(\mu_0 l_0^2 D)$ to be large (in the range $1 \ll Pe \ll l_0^2/h_0^2$), ensuring that in-plane diffusion is weak, {\color{black} but sufficient to suppress gradients across the thickness of the film.  (In practice, persistent mucin gradients across the short dimension of the film are a robust feature of airway liquid; for a computational treatment of such effects see \cite{chatelin2016}.)}

We do not model the collision of a droplet with the layer, but instead suppose that a droplet has an initially axisymmetric parabolic shape and sits directly on a precursor film of thickness $\eta \ll 1$ with
\begin{align}
H(x,y,0) = \begin{cases}
\eta + 1 - x^2 -y^2, & (x^2+y^2<1), \\
\eta, & (x^2+y^2\geq1).
\end{cases}
\label{eq:ic}
\end{align}
The initial viscosity field $M(x,y,0)$ is represented as a log-normal random field $\mathcal{M} = \exp{(\sigma \mathcal{G})}$, where $\sigma$ is a prescribed amplitude and $\mathcal{G}$ is a stationary Gaussian random field with zero mean and squared exponential covariance
\begin{align}
k_\mathcal{G}(x, x';y,y') = \exp\left(-\frac{(x-x')^2+(y-y')^2}{2\ell^2}\right),
\label{eq:covar}
\end{align}
where $\ell$ is a prescribed correlation length. 

When $\sigma = 0$, the initial viscosity $M(x,y,0) \equiv 1$ is homogeneous and the drop spreads axisymmetrically. When $\sigma$ is nonzero, the initially axisymmetric drop is distorted by the heterogeneous initial viscosity as it spreads. % (see figure~\ref{single_simulation}). 
We characterise the motion using the area and the position of the centre of mass of the distorted drop,
\begin{align}\label{qoi_definition}
A(t) = \int_{\mathcal{A}} \mathrm{d}A, \quad \bm{D}(t) &= \frac{1}{A}\int_{\mathcal{A}} (x\bm{i} +y \bm{j})\mathrm{d}A \equiv D_x \bm{i} + D_y \bm{j},
\end{align}
where $\mathcal{A}$ is the domain under the drop, and $\bm{i}$ and $\bm{j}$ the unit vectors in the $x$ and $y$ directions. The uncertainties of the initial random viscosity field, parametrised by $\sigma$ and $\ell$, propagate through (\ref{2d_vector_form}) to give outputs at a fixed time $t_f$ as random variables $A(t_f)$, $D_x(t_f)$, $D_y(t_f)$ and $D(t_f) \equiv \sqrt{D_x^2+D_y^2}$. In the following we investigate how these outputs, particularly the variances $\sigma^2_A$, $\sigma^2_{D_x}$ and $\sigma^2_{D_y}$, depend on the input random field.

{\color{black}Details of the numerical method for solving (\ref{2d_vector_form},\ref{eq:ic}), and of a GP emulator of the model, allowing efficient estimation of the parameter-dependence of the variances $\sigma_A^2$ and $\sigma_{D_y}^2$, are given in Appendix~\ref{app:sim} and \ref{sec:gp} respectively.}

\section{Numerical results}
\label{sec:sim}

\begin{figure}
\begin{center}
\begin{overpic}[width = 0.45\linewidth]{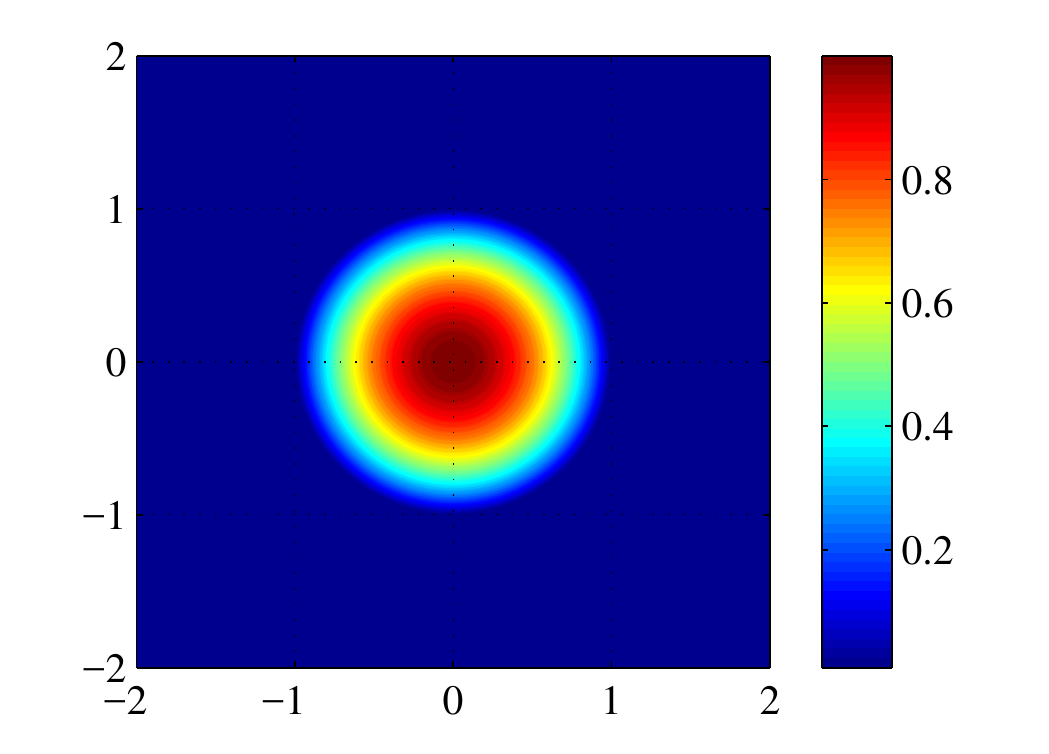}
\put(0,60) {(a)}
\put(4,37) {$y$}
\put(42,-2) {$x$}
\end{overpic}
\begin{overpic}[width = 0.45\linewidth]{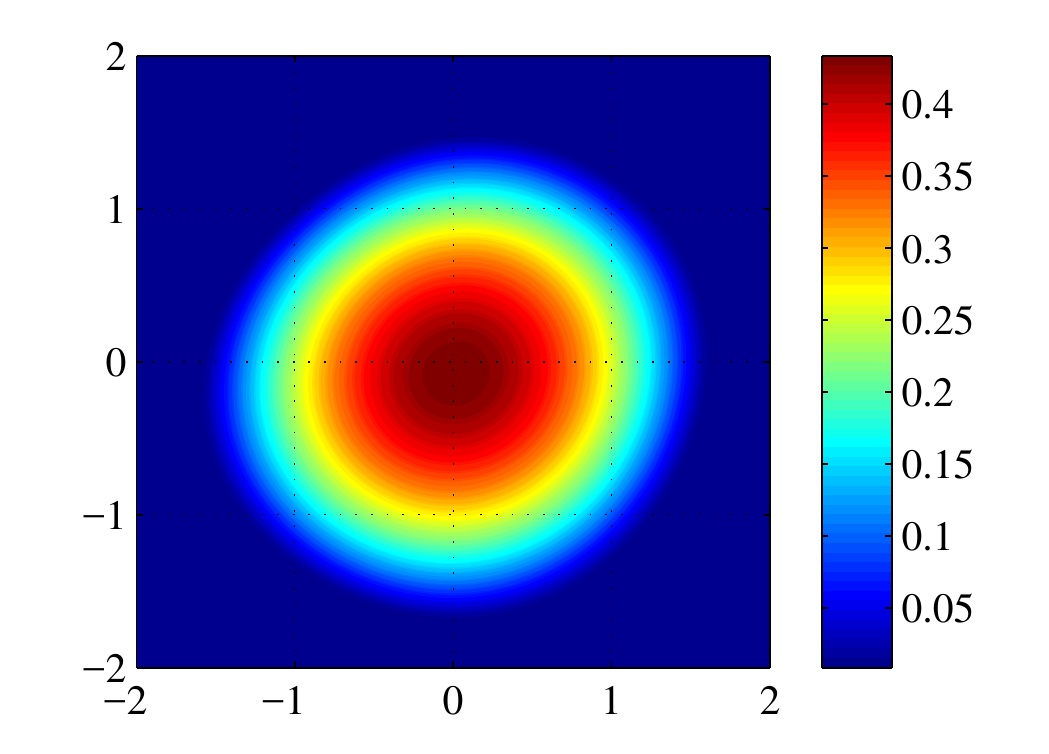}
\put(0,60) {(b)}
\put(4,37) {$y$}
\put(42,-2) {$x$}
\end{overpic}
\vskip 0.25in
\begin{overpic}[width = 0.45\linewidth]{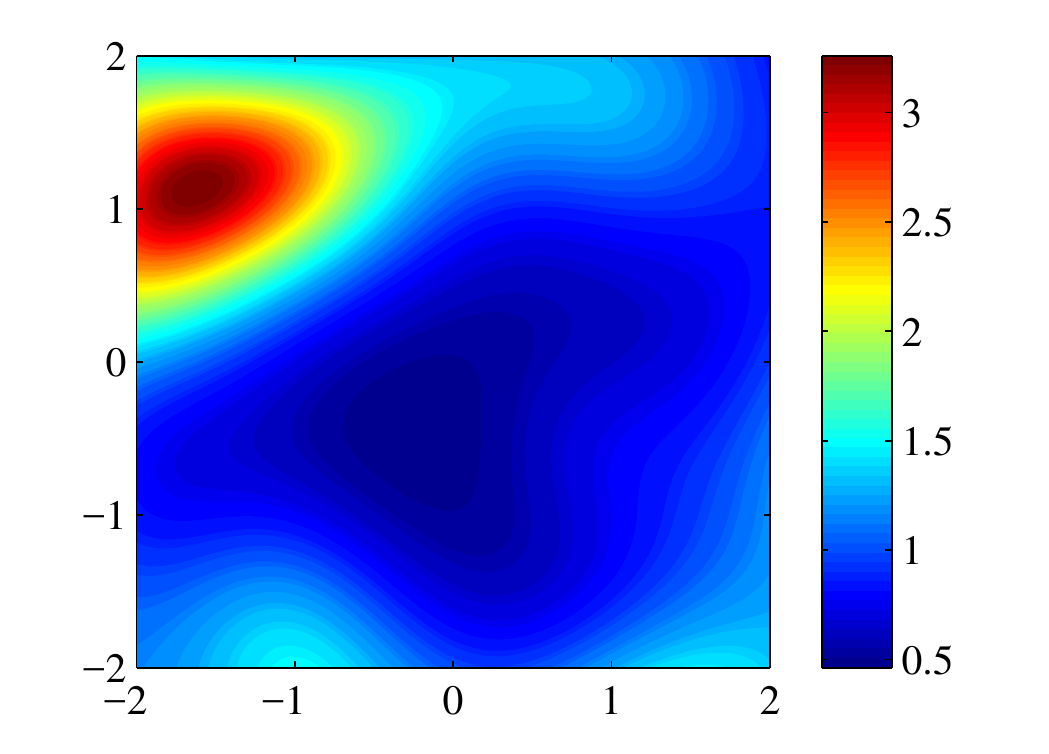}
\put(0,60) {(c)}
\put(4,37) {$y$}
\put(42,-2) {$x$}
\end{overpic}
\begin{overpic}[width = 0.45\linewidth]{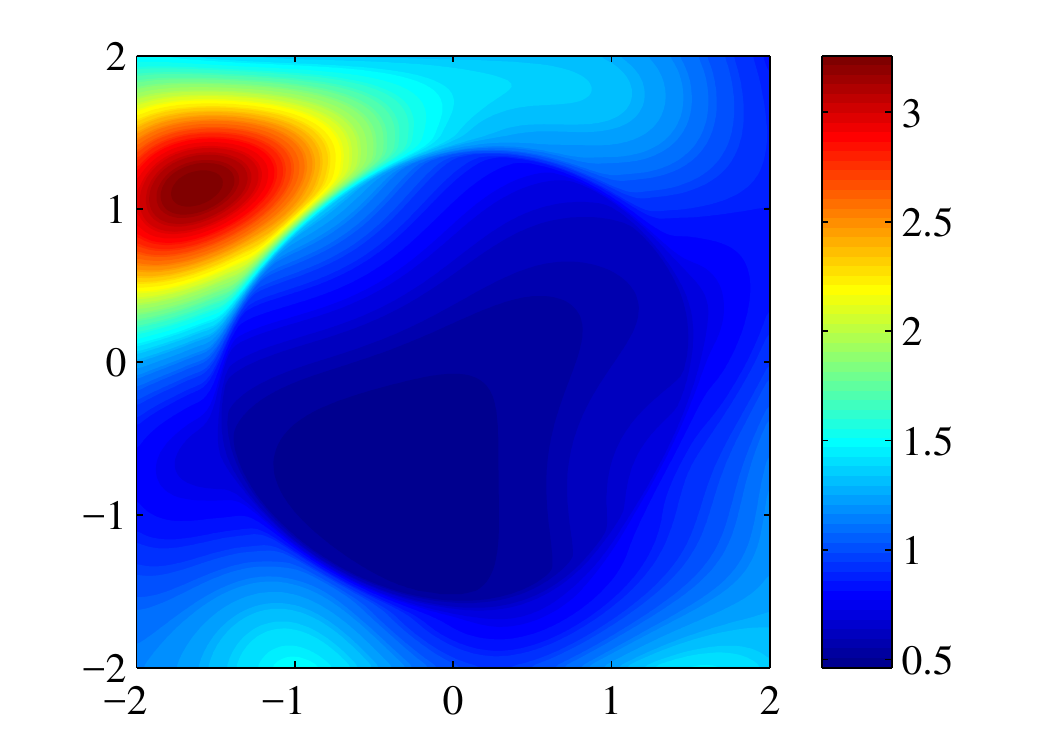}
\put(0,60) {(d)}
\put(4,37) {$y$}
\put(42,-2) {$x$}
\end{overpic}
\vskip 0.25in
\begin{overpic}[width = 0.45\linewidth]{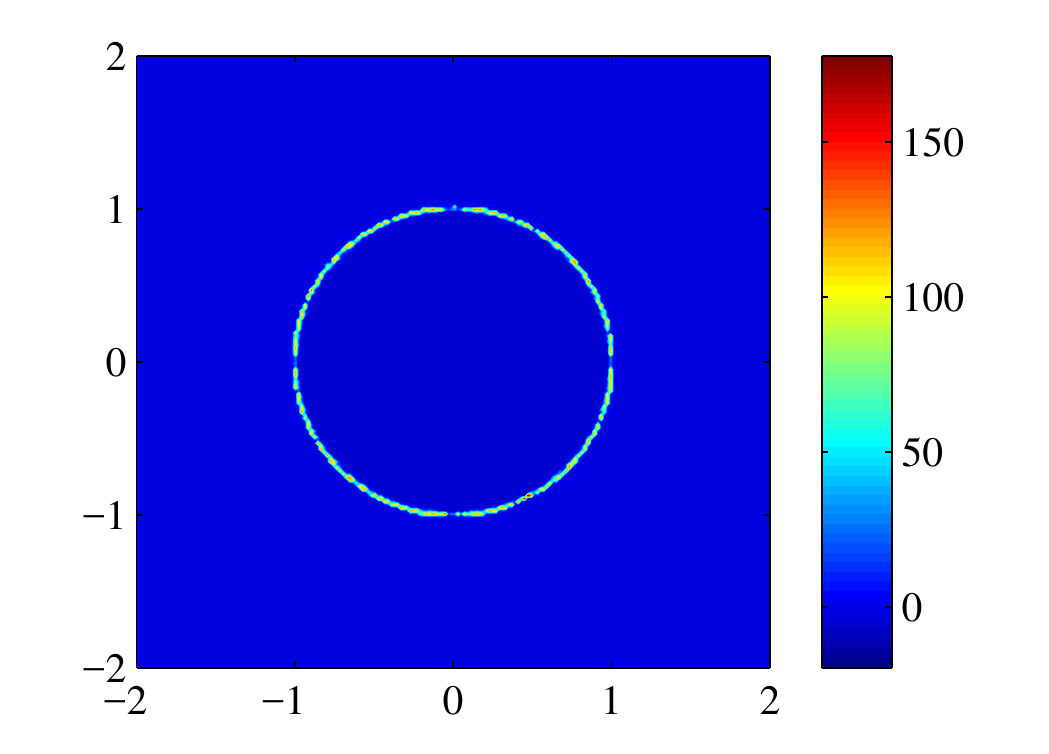}
\put(0,60) {(e)}
\put(4,37) {$y$}
\put(42,-2) {$x$}
\end{overpic}
\begin{overpic}[width = 0.45\linewidth]{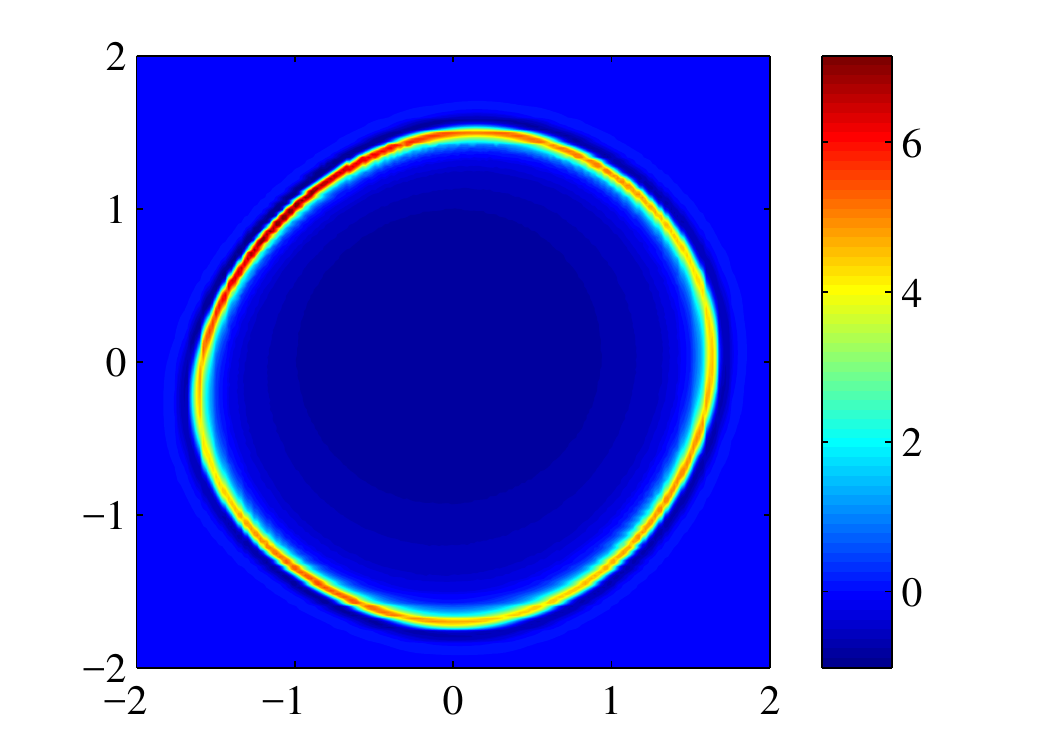}
\put(0,60) {(f)}
\put(4,37) {$y$}
\put(42,-2) {$x$}
\end{overpic}
\caption{Numerical simulations of (\ref{2d_vector_form}) with the initial viscosity field generated from the random field $\mathcal{M}$ with $\sigma = 0.5$ and $\ell = 1$: (a) $H(x,y,0)$, (b) $H(x,y,10)$, (c) $M(x,y,0)$, (d) $M(x,y,10)$, (e) $\nabla^2H(x,y,0)$, (f) $\nabla^2H(x,y,10)$; remaining parameters are $\eta = 0.01$ and $Pe = 10^5$.}
\label{single_simulation}
\end{center}
\end{figure}

\begin{figure}
\begin{center}
%\begin{overpic}[width = 0.45\linewidth]{Area_evolution.eps}
\begin{overpic}[width = 0.45\linewidth]{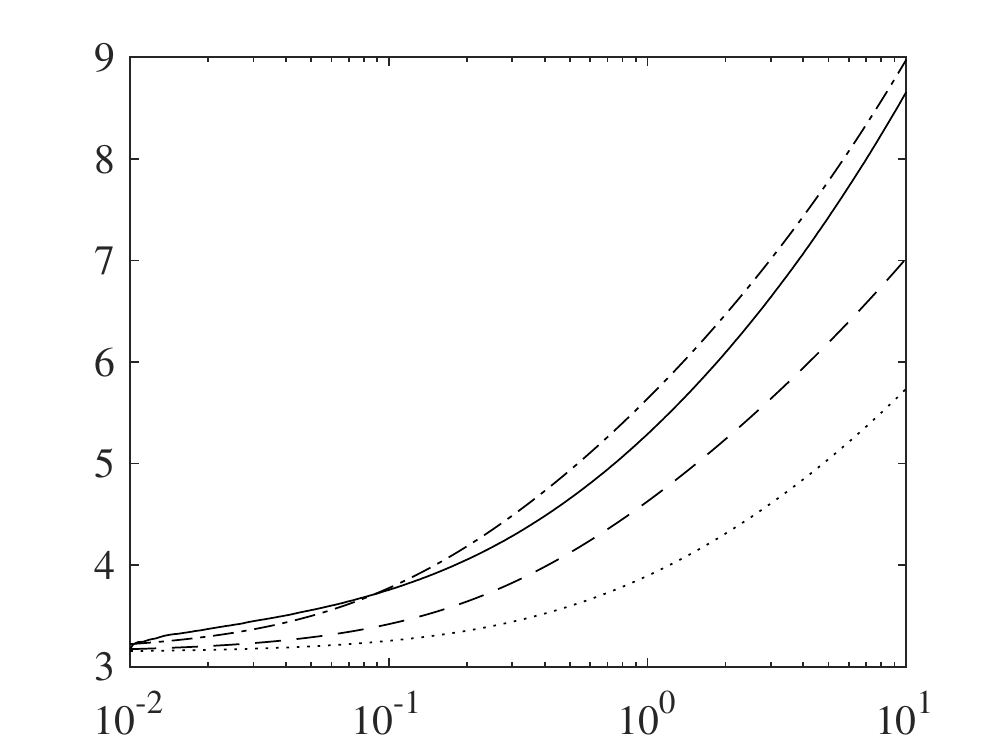}
\put(15,60) {(a)}
\put(50,-5) {$t$}
\put(-2,40) {$A$}
\end{overpic}
%\begin{overpic}[width = 0.45\linewidth]{CoMx_evolution.eps}
%\put(15,60) {(b)}
%\put(50,-5) {$t$}
%\put(-4,40) {$D_x$}
%\end{overpic}
%\vskip 0.25in
%\begin{overpic}[width = 0.45\linewidth]{CoMy_evolution.eps}
%\put(15,20) {(c)}
%\put(50,-5) {$t$}
%\put(-4,40) {$D_y$}
%\end{overpic}
\begin{overpic}[width = 0.45\linewidth]{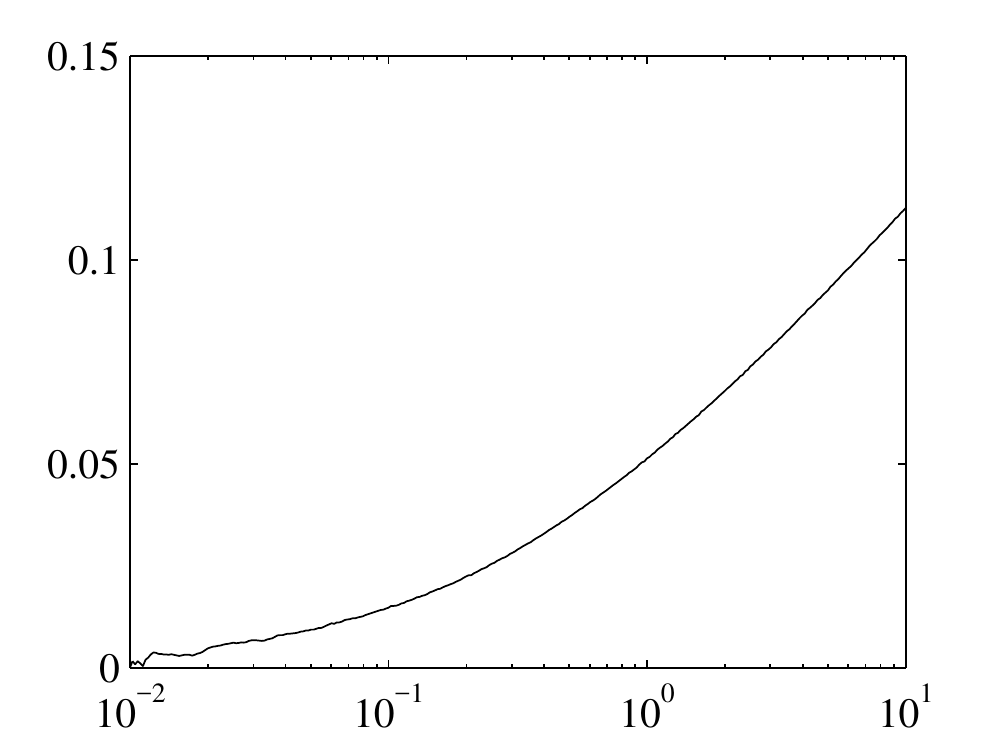}
\put(13.5,28.6) {\includegraphics[width = 0.225\linewidth]{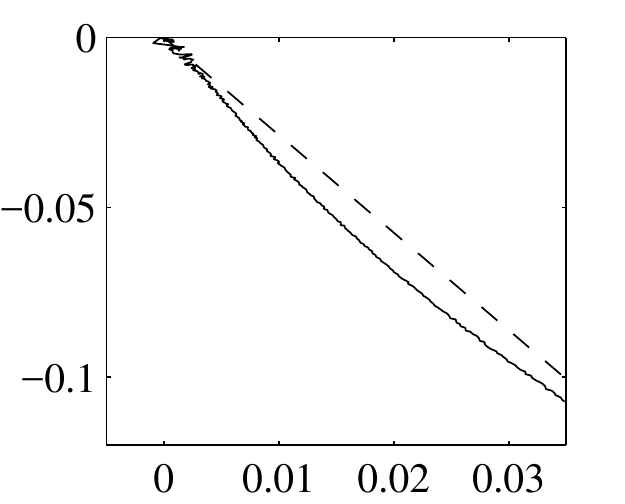}}
\put(50,-5) {$t$}
\put(-4,40) {$D$}
\put(80,20) {(b)}
\put(60,50) {$D_y$}
\put(37,24) {$D_x$}
\end{overpic}
\caption{Time evolution of (a) $A(t)$ and (b) $D(t)$, for {\color{black}a droplet spreading on a heterogeneous film in the realisation shown} in figure~\ref{single_simulation}. {\color{black} (a) shows simulation (solid), the leading-order asymptotic approximation $\pi a_0^2$ using (\ref{LOCL}) for $a_0$ (dotted), the leading-order approximation plus the correction accounting for heterogeneity (\ref{eq:areaX}) with $b_0$ satisfying (\ref{LeadingPerturbation}) (dashed), and the latter using instead (\ref{eq:a0refined}) for $a_0$, accounting for the pressure gradient across the bulk of the drop (dot-dashed)}. The inset in (b) shows $D_x$ and $D_y$, indicating the path of the drop centre of mass, showing simulations (solid) and the asymptotic approximation  (\ref{FirstPerturbationCos}, \ref{FirstPerturbationSin}) (dashed),  using (\ref{LOCL}) for $a_0$.  
% In this example $\eta=0.01$ ($1/\log(1/\eta)\approx 0.217$).
}
\label{single_simulation_evolution}
\end{center}
\end{figure}

Figure~\ref{single_simulation} shows an example of a drop spreading over a thin strongly heterogeneous liquid layer (with $\sigma=0.5$).  In this example, the initially axisymmetric drop (Figure~\ref{single_simulation}a) spreads towards a region of elevated viscosity located in the upper left domain (Figure~\ref{single_simulation}c).  As the drop spreads, the pressure field $-\nabla^2 H$ remains almost uniform everywhere except for a narrow boundary layer at the advancing contact line (Figure~\ref{single_simulation}e,f), where the pressure has a deep local minimum.  At later times, the drop shape becomes distorted (Figure~\ref{single_simulation}b,f), with spreading hindered locally by the higher viscosity fluid.  The viscosity field is distorted by the spreading (Figure~\ref{single_simulation}d), being stretched beneath the drop but with large gradients developing across the contact line.  The drop area increases with time (Figure~\ref{single_simulation_evolution}a, {\color{black}solid line}) and the drop's centre of mass drifts away from the origin towards the lower right domain (Figure~\ref{single_simulation_evolution}b, {\color{black}solid line}).  {\color{black}(Broken lines in Figure~\ref{single_simulation_evolution} show predictions of the asymptotic model which we shall return to in Sec.~\ref{sec:3} below.)}

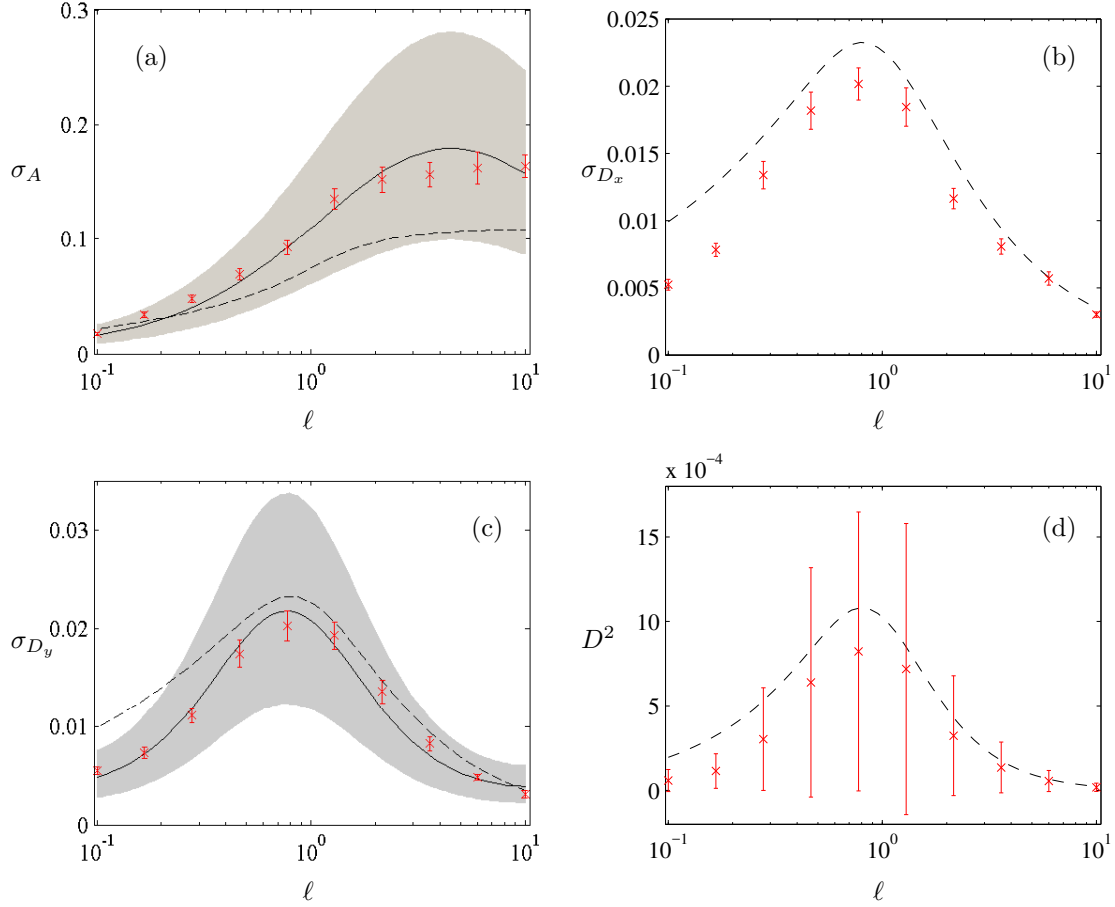
\begin{figure}
\begin{center}
\begin{overpic}[width=0.45\linewidth]{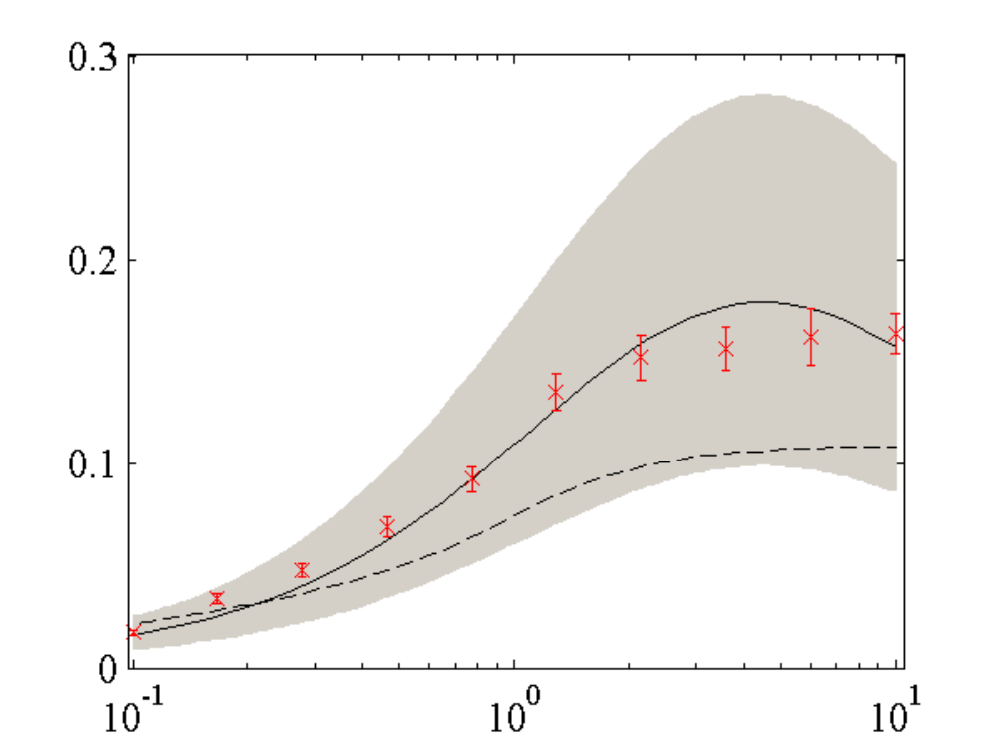}
\put(20,60) {(a)}
\put(50,-5) {$\ell$}
\put(-2,40) {$\sigma_A$}
\end{overpic}
\begin{overpic}[width=0.45\linewidth]{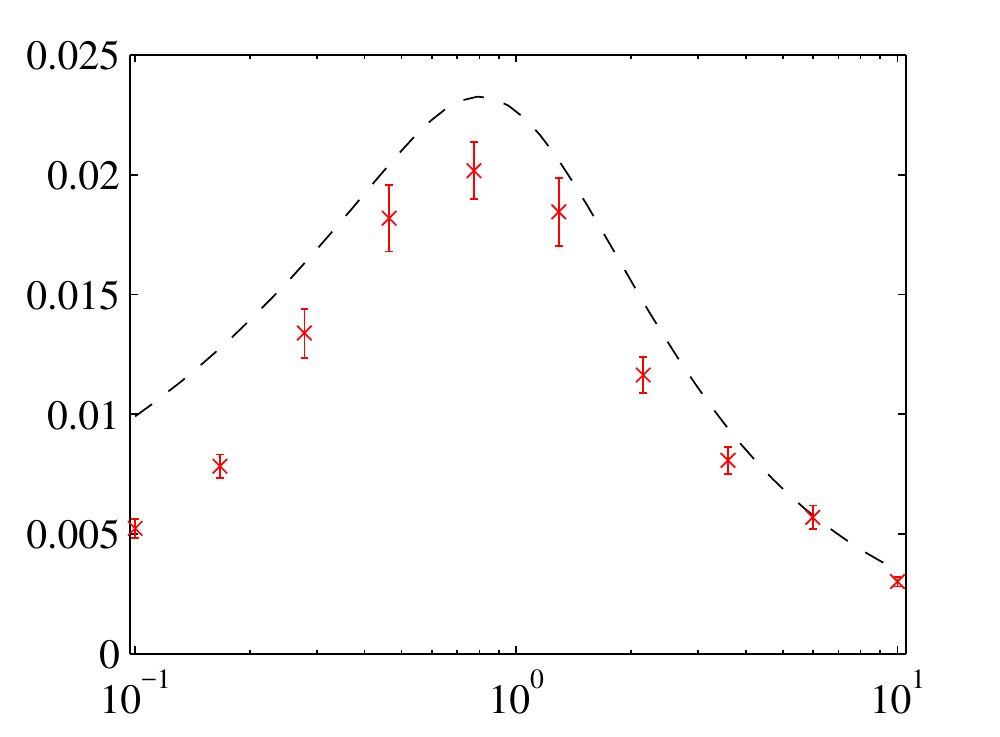}
\put(80,60) {(b)}
\put(50,-5) {$\ell$}
\put(-2,40) {$\sigma_{D_x}$}
\end{overpic}
\vskip 0.25in
\begin{overpic}[width=0.45\linewidth]{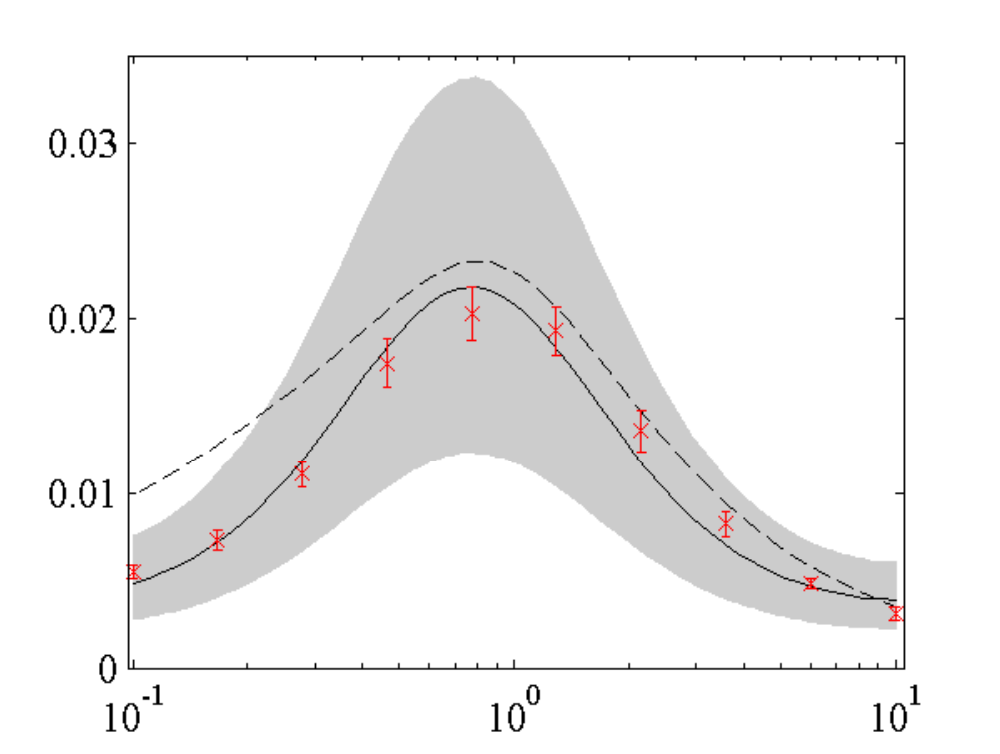}
\put(80,60) {(c)}
\put(50,-5) {$\ell$}
\put(-2,40) {$\sigma_{D_y}$}
\end{overpic}
\begin{overpic}[width=0.45\linewidth]{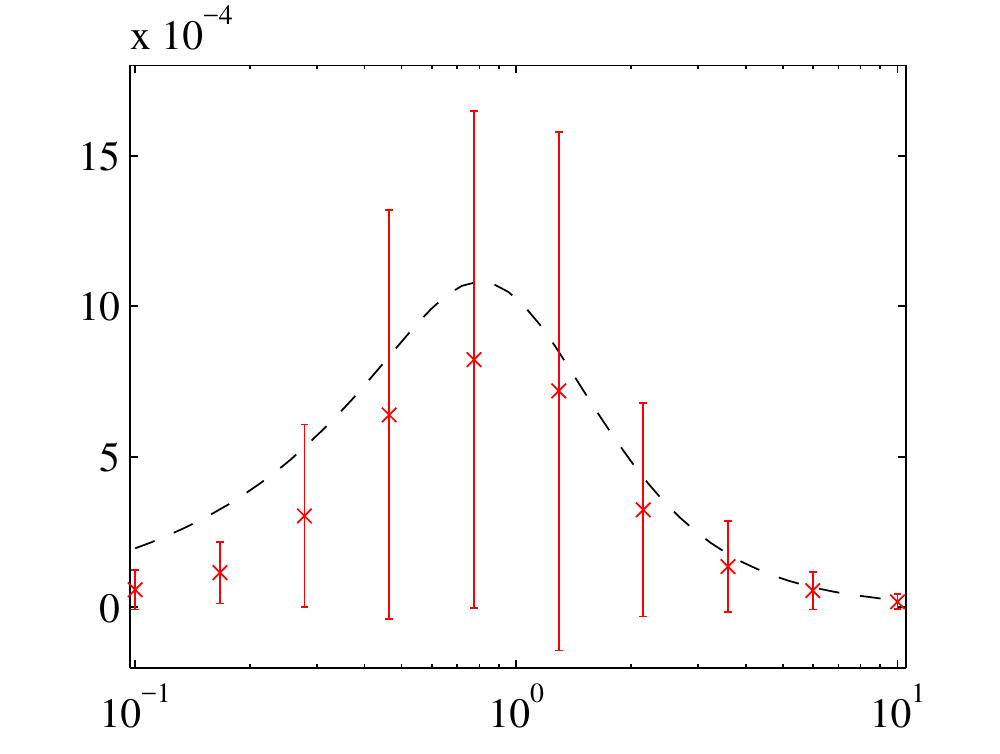}
\put(80,60) {(d)}
\put(50,-5) {$\ell$}
\put(-2,40) {$D^2$}
\end{overpic}
\caption{The effect of correlation length $\ell$ on (a) the standard deviation of the drop area $A(t_f)$, (b, c) the $x$ and $y$-coordinate of the centre of mass $D_x(t_f)$ and $D_y(t_f)$, and (d) the mean and one standard deviation error of the square radial displacement $D^2(t_f)$ at $t_f = 10$. The solid lines in (a) and (c) represent predicted mean estimates from emulation results, with shading in (a) and (c) showing 95\% confidence intervals% (broad after transformation from intervals for $\log(\mathrm{Var})$)
; the dashed lines represent asymptotic predictions (\ref{VarA}, \ref{VarDx}, \ref{VarDy}, \ref{MeanR2}); red error bars in (a - c) show bootstrapping results calculated from MC simulations of (\ref{COMSOL}); red error bars in (d) show MC estimates of standard deviation. {\color{black} $D^2$ has an asymmetric distribution about its mean because of the constraint $D^2\geq 0$}.  Parameters are $\sigma = 0.1$, $\eta = 0.01$ and $Pe = 10^{5}$.}
\label{errorbar}
\end{center}
\end{figure}

Conducting a large set of such simulations, each involving a new realisation of the initial viscosity field, provides us with a dataset from which we can estimate the variability in drop size and location as a function of the input parameters.  Symbols with bars in Figure~\ref{errorbar} show results for $\sigma=0.1$ at ten values of the correlation length $\ell$ of the viscosity field.  The variance of drop area (Figure~\ref{errorbar}a, {\color{black}symbols with bars}), determined pointwise by MC simulation, increases with $\ell$ over the range presented.  As seen in previous studies \citep{savva2010, 2016-Xu-vol472}, short-range fluctuations are uncorrelated across the drop whereas long-range variations lead to more systematic deviations from the mean (the contact line's advance is uniformly helped or hindered by correlated reductions or increases in the $M$ field, leading to greater variability in drop area).  The mean-square lateral drift $\mathbb{E}[D^2]$ is greatest when the correlation length is comparable to the initial drop diameter (Figure~\ref{errorbar}d, {\color{black}symbols with bars}): small $\ell$ leads to cancellation of uncorrelated fluctuations and large $\ell$ reduces gradients across the drop that might displace it in a preferred direction.  The variability of the drift increases in proportion to the mean.  This is illustrated also by the variance of the horizontal and vertical drift (Figure~\ref{errorbar}b,c, {\color{black}symbols with bars}), which are again greatest for $\ell$ close to unity.  

The pointwise variance estimates presented in Figure~\ref{errorbar}(a--c) were obtained from $100$ simulations at 10 different values of the input parameters $\ell$ with fixed $\sigma = 0.1$.  Bootstrapping was used to estimate the sampling error in each case.  To estimate variances at intermediate parameters, or over a wider region of $(\sigma,\ell)$-space, a more efficient method is required that makes better use of limited simulation data.  To achieve this we emulated the outcome of simulations as a GP (see Appendix~\ref{sec:gp}), running 500 PDE simulations at parameter values determined via an optimised Latin Hypercube (oLHC) design algorithm (sampling over $\log_{10} \ell\in (-1.1,1.1)$ with $\sigma=0.1$).  Variance predictions based on 250 simulations from this training dataset (one of a dozen random samples from the 500 simulations, of which this is representative) are shown in Figure~\ref{errorbar}(a,c) as a mean (solid) with 95\% confidence intervals (shaded).  The mean estimated variance agrees well with the pointwise MC estimates, although the maximum in the estimated mean variance $\sigma_A$ is not evident in the MC data.   The maximum in $\sigma_{D_y}$ is well captured, however.  (We shall discuss asymptotic predictions, represented by dashed lines in Figure~\ref{errorbar}, in Sec.~\ref{sec:3} below.)

\begin{figure}
	\begin{center}
	\begin{overpic}[width=0.45\linewidth]{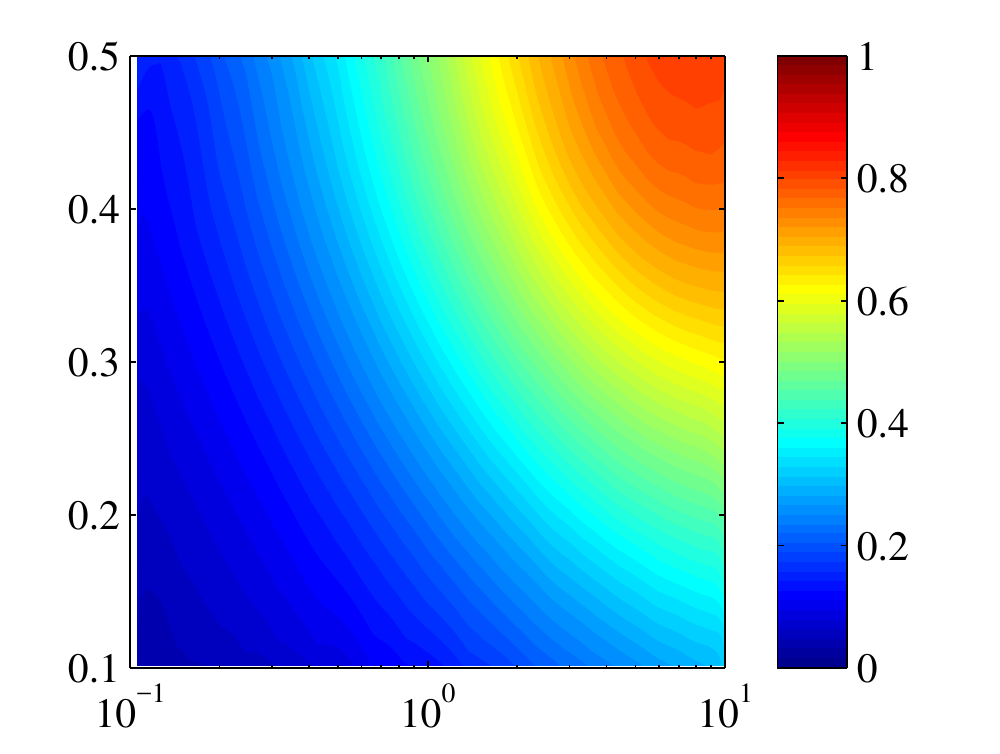}
    \put(0,65) {(a)}
	\put(42,-3) {$\ell$}
	\put(-3,40) {$\sigma$}
    \put(80,75) {$\sigma_A$}
	\end{overpic}
	\begin{overpic}[width=0.45\linewidth]{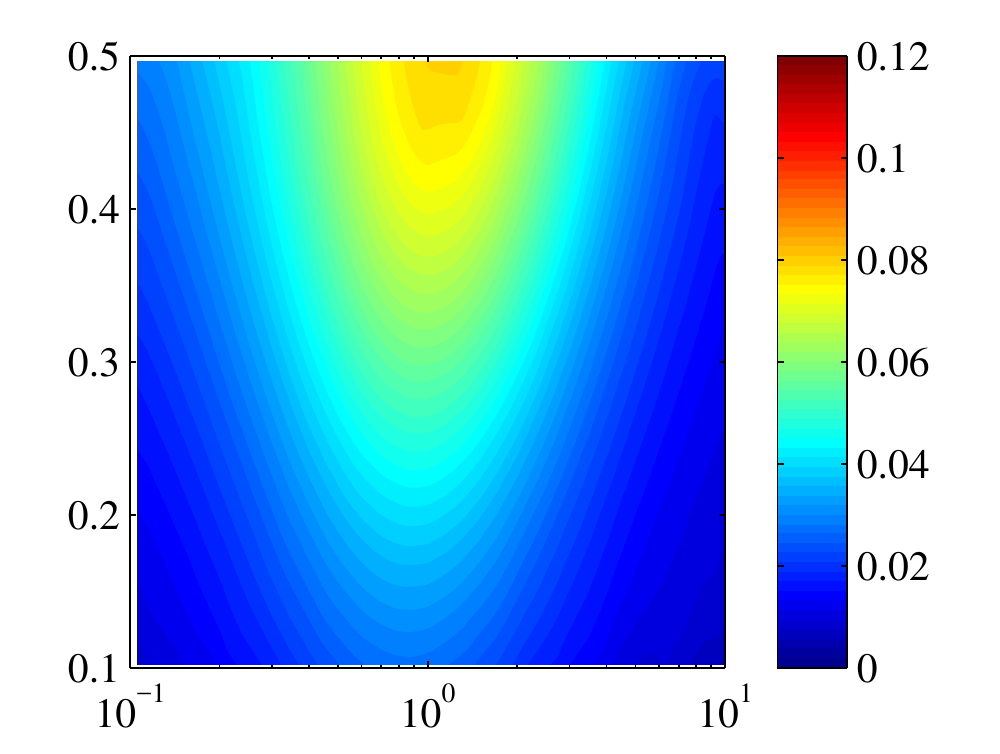}
    \put(0,65) {(b)}
	\put(42,-3) {$\ell$}
	\put(-3,40) {$\sigma$}
    \put(80,75) {$\sigma_{D_y}$}
	\end{overpic}
	\vskip 0.25in
	\begin{overpic}[width=0.45\linewidth]{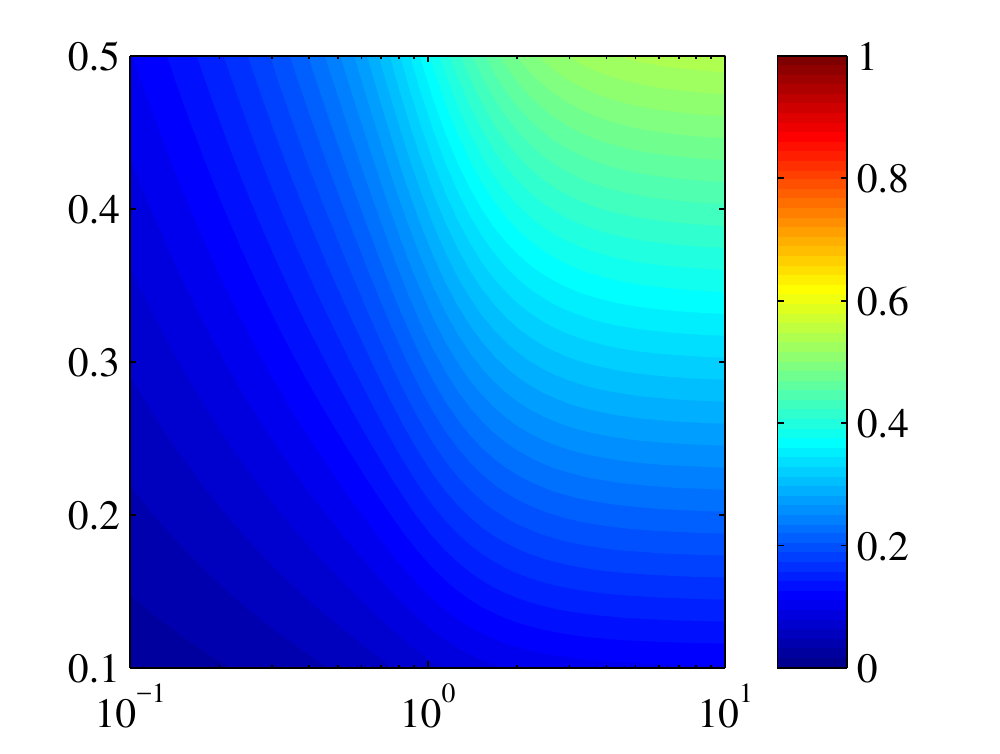}
    \put(0,65) {(c)}
	\put(42,-3) {$\ell$}
	\put(-3,40) {$\sigma$}
    \put(80,75) {$\sigma_A$}
	\end{overpic}
	\begin{overpic}[width=0.45\linewidth]{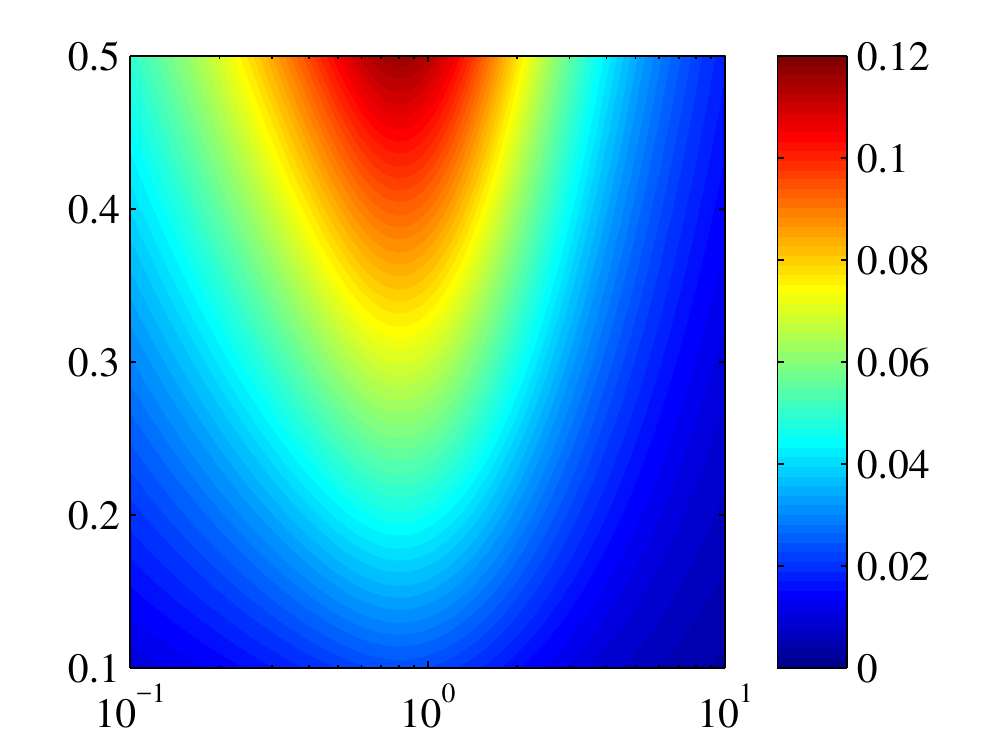}
    \put(0,65) {(d)}
	\put(42,-3) {$\ell$}
	\put(-3,40) {$\sigma$}
    \put(80,75) {$\sigma_{D_y}$}
	\end{overpic}
	\caption{The dependence of $\sigma_A$ and $\sigma_{D_y}$ on the standard deviation $\sigma$ and the correlation length $\ell$ of the input random field.  (a, b) show emulation estimates; (c, d) show asymptotic predictions (\ref{VarA}, \ref{VarDy}).  Remaining parameters are $\eta = 0.01$ and $Pe = 10^5$.}
	\label{fig:emuls2D}
	\end{center}
\end{figure}
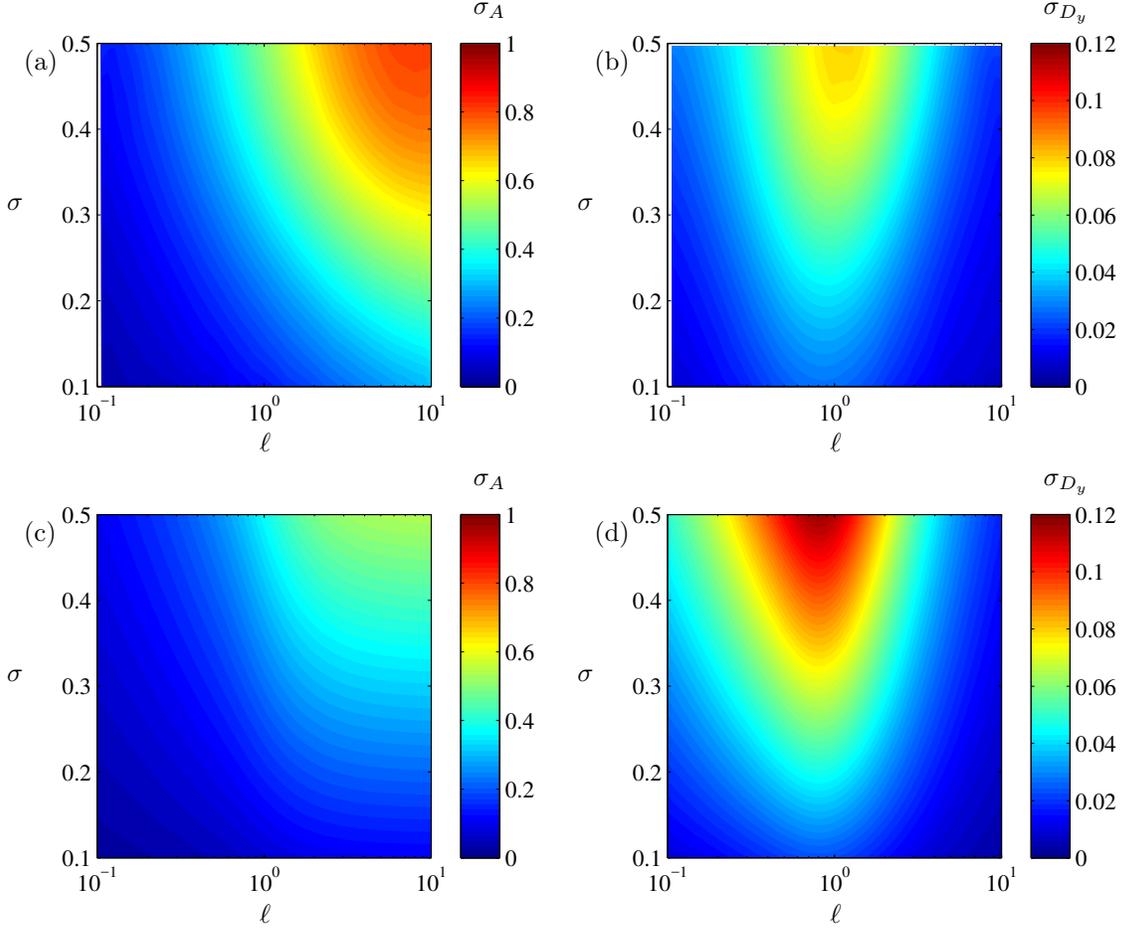

We then extended variance predictions over the domain $0.1\leq \sigma\leq 0.5$, $0.1\leq \ell \leq 10$ using a set of 250 simulations over the domain (12 such sets were selected randomly from a set of 500 simulations in all, specified using an oLHC design over $(\log_{10}(\ell),\sigma)$-space).  Results are shown in Figure~\ref{fig:emuls2D}(a,b) for the drop area and vertical drift.  Simulations for larger $\sigma$ were challenging because of the emergence of very large gradients in the solute viscosity field {\color{black}near the drop's effective contact line}; the 12 sample sets were chosen to ensure that the predictions in Figure~\ref{fig:emuls2D}(a,b) were robust.  The wider sample leads to an estimated mean $\sigma_A$ that is monotonic in both $\ell$ and $\sigma$ (Figure~\ref{fig:emuls2D}a).  The estimated mean variance of vertical displacement retains a maximum for $\ell$ close to unity (for fixed $\sigma$), which increases with respect to $\sigma$ (Figure~\ref{fig:emuls2D}b).  A key benefit of the GP emulation over pointwise MC estimates is that information about the variance at each input parameter value (collected over an input-space filling design) contributes to knowledge about the variance at other input values, making estimates over a wide range of parameter space accessible.  

Figures~\ref{single_simulation_evolution}, \ref{errorbar} and \ref{fig:emuls2D}(c,d) also show asymptotic predictions of variances, which we now derive; the predictions are evaluated in Sec.~\ref{sec:comp} below.

\section{Weak-disorder asymptotics}
\label{sec:3}

We develop an approximation in the limit $\eta \ll 1$ (ensuring that the flow near the {\color{black}drop's effective} contact line determines spreading rates), $\sigma \ll 1$ (implying that random effects appear as a linear perturbation to a deterministic axisymmetric flow) and $Pe\gg 1$ (implying that the viscosity field is transported along characteristics of the spreading flow).  The effects of weak in-plane diffusion become significant only at large times \citep{2016-Xu-vol472}; we restrict attention here to early to moderate times and henceforth neglect diffusion.  Our aim is to derive explicit leading-order predictions for properties of the random variables in Figures~\ref{single_simulation_evolution}--\ref{fig:emuls2D}.  {\color{black}With this in mind, we use the simplest model that captures the effects of interest, trading some accuracy for clarity.}

\subsection{The axisymmetric base state}

When the precursor layer is thin, the spreading is slow and the drop is almost in equilibrium.  It is convenient to define $\epsilon\equiv 1/\log(1/\eta)$ and set $t=T/\epsilon$, $\bm{u}=\epsilon \mathbf{U}$.  {\color{black}(This slow timescale for spreading was identified for example by \cite{hocking1983} and \cite{king2001}.)}  Our starting point from (\ref{2d_vector_form}) is therefore 
\begin{equation}
H_T+\nabla\cdot(H\mathbf{U})=0,\quad M_T+\mathbf{U}\cdot\nabla M=0, \quad \epsilon \mathbf{U}=\frac{H^2}{3M}\nabla \nabla^2 H.
\label{eq:hmu}
\end{equation}
The base state in the limit $\sigma\rightarrow 0$ has $M=1$ and $H=H(r,T)$, {\color{black} where $r$ is the radial polar coordinate}.  The leading-order solution of (\ref{eq:hmu}) as $\epsilon\rightarrow 0$ can then be decomposed into an outer region (in which $H$ has uniform curvature) and an inner region near the drop's contact line governed by the Landau--Levich equation.  In the outer region, in $0\leq r<a(T)$,
\begin{equation}
H=\frac{2V}{\pi a^2}\left(1-\frac{r^2}{a^2}\right),\quad \mathbf{U}=\frac{a_T }{a}r \hat{\mathbf{r}}
\label{eq:axisol}
\end{equation}
where the drop volume is $V=\pi/2$, matching the initial condition (\ref{eq:ic}).  (The next-order in $\epsilon$ correction to this approximation is given in Appendix C.)  The drop's contact angle is $\Theta=-H_r\vert_{r=a-}=4V/(\pi a^3)$.  In $r>a$, the film is undisturbed with $H=\eta$ uniformly.  The inner region connecting these states is defined by setting 
\begin{equation}
\label{eq:innerscale}
r=a+(\eta/(\epsilon a_T)^{1/3})X\quad\mathrm{and}\quad H=\eta G(X).  
\end{equation}
Imposing $G\rightarrow 1$ for $X \rightarrow \infty$, (\ref{eq:hmu}) reduces to $\tfrac{1}{3}G^3 G_{XXX}=G-1$ \citep{hocking1982,2016-Xu-vol472}.  This well-known ODE has a unique solution (up to translation) for which $G_{XX}\rightarrow 0$ as $X\rightarrow -\infty$.  This has the outer limit $G\sim - 9^{1/3}X \log\vert X\vert+\dots$ as $X\rightarrow -\infty$.  When expressed in outer variables, the drop's contact angle (at the outer limit of the inner solution) satisfies \citep{2016-Xu-vol472}
\begin{equation}
\label{CubicInner}
 \Theta^3 = -H_r^3 = 9 a_T \left(1+ \epsilon\left( \ln|a - r| +\tfrac{1}{3}\ln|3a_T | +\tfrac{1}{3}\ln \epsilon + 1 + 3^{2/3} \alpha\right)+\dots \right) 
\end{equation}
where $\alpha$ is an order unity constant.  Matching expressions for $\Theta^3$ between the inner and outer solutions gives the leading-order drop spreading-rate as 
\begin{equation}
\label{eq:a0}
9a_T=(4V/\pi a^3)^3.
\end{equation}
Thus
\begin{align}\label{LOCL}
a = \frac{(640 T V^3+9 \pi ^3)^{1/10}}{3^{1/5} \pi ^{3/10}} = \left(1+\frac{80 T}{9}\right)^{1/10} \equiv \left(1+\frac{80 t}{9 \ln(1/\eta)}\right)^{1/10},
\end{align}
{\color{black}using} the initial condition $a_0(0) = 1$ and $V = \pi/2$.  The corrections to this approximation, appearing at $O(\epsilon)$ {\color{black}and given in (\ref{eq:a0refined}) below}, can be appreciable unless $\eta$ is vanishingly small {\color{black}(as illustrated in Figure~\ref{fig:axisym} in Appendix~\ref{sec:hior} below)}, and have been examined in detail previously \citep{hocking1982,savva2009, sibley2015,2016-Xu-vol472}.  They require evaluation of pressure gradients across the bulk of the drop, which are just visible in Figure~\ref{single_simulation}(f).  Here, however, we will restrict attention to the leading-order approximation as it makes the study of variable viscosity and non-axisymmetric spreading more transparent {\color{black}and is sufficient to reveal the dominant physical processes}.  In the process, we accept a loss in quantitative accuracy whenever $\eta$ is not exponentially small.

We now consider perturbations of this axisymmetric solution in the formal limit 
\begin{equation}
\eta\ll \eta/\epsilon^{1/3}\ll \epsilon \ll \sigma \ll 1\quad{\color{black}\mathrm{for}\quad T=O(1).}
\label{eq:limits}
\end{equation}
The first inequality ensures that the precursor layer thickness is much thinner than the width of the inner region at the contact line; the second and third ensure that the inner region is thin in comparison to perturbations in the shape of the contact line (as in Figure \ref{single_simulation}f); the third ensures that the perturbations due to disorder do not disrupt the capillary equilibrium structure in the main drop; the fourth ensures that the drop does not deviate significantly from the axisymmetric base state.  {\color{black}In addition, the condition $t\ll (\log(1/\eta))/\eta^5$ ensures that the drop height remains large in comparison to the precursor film thickness (estimating $H(0,T)\propto T^{-1/5}\gg \eta$ for $T\gg 1$ from (\ref{eq:axisol}, \ref{LOCL})).  Although the condition (\ref{eq:limits}) is restrictive, the associated approximation reveals features of the flow that are likely to be relevant to a wider region of parameter space.}

\subsection{The non-axisymmetric outer problem}

In the outer region describing the main body of the drop, we expand using
\begin{align}
\label{weak_disorder_expansions}
H(r,\theta,T) = H_0(r,T) + \sigma H_1(r,\theta,T) + O(\sigma^2), \quad M = 1 + \sigma M_1 + O(\sigma^2), \quad \mathbf{U} = \mathbf{U}_0 + \sigma \mathbf{U}_1 + O(\sigma^2),
\end{align}
where $H_0$ and $\mathbf{U}_0$ are the axisymmetric solution given in (\ref{eq:axisol}) {\color{black} and $\theta$ is the polar angle}.  The contact line location is also expanded as
\begin{equation}
\label{eq:aexp}
a(\theta,T)=a_0(T)+\sigma a_1(\theta,T)+O(\sigma^2)
\end{equation}
where $a_0$ satisfies (\ref{eq:a0}, \ref{LOCL}).  Substituting (\ref{weak_disorder_expansions}) into (\ref{eq:hmu}) and expanding gives, at different orders of $\sigma$,
\begin{subequations}
\label{SigmaEquations}
\begin{align}
O(1): & H_{0T} + \nabla \cdot (H_0 \bm{U}_0) = 0, \quad \epsilon\bm{U}_0 = \frac{H_0^2}{3} \nabla \nabla^2 H_0, \\
O(\sigma): & H_{1T} + \nabla \cdot (H_0 \bm{U}_1 + H_1 \bm{U}_0) = 0, \quad M_{1T} + \bm{U}_0 \cdot \nabla M_1 = 0, \nonumber \\
& \epsilon \bm{U}_1 = \frac{H_0}{3}\left(2 H_1 \nabla \nabla^2 H_0 - H_0 M_1 \nabla \nabla^2 H_0 + H_0 \nabla \nabla^2 H_1 \right).
\end{align}
\end{subequations}
The constraint (\ref{eq:limits}) ensures that viscous terms (proportional to $\epsilon$) can be neglected in (\ref{SigmaEquations}), ensuring $\nabla \nabla^2 H_0=0$ and $\nabla \nabla^2 H_1=0$.  We impose $H_1\rightarrow 0$ as $r\rightarrow a-$, and a volume constraint on $H_1$, to give 
\begin{equation}
\label{eq:h1}
\nabla^2 H_1=P_1(T)\quad (r<a_0),\quad
\lim_{r \to a_0-} (a_1 H_{0r} + H_1) = 0, \quad \int_0^{2\pi} \int_0^{a_0} r H_1 \mathrm{d}r \mathrm{d}\theta = 0.
\end{equation}
$P_1$ can be chosen to enforce the volume constraint.  Meanwhile the non-uniform viscosity field $M_1$ is transported radially following (\ref{SigmaEquations}b), remaining constant along characteristic curves $r = r_c(T)$ satisfying
% \begin{align}
% \label{CharCurveEq}
${\mathrm{d} r_c}/{\mathrm{d} T} = U_0$.
% \end{align}
Using the expression for $\bm{U}_0$ in (\ref{eq:axisol}), the characteristic curves can be expressed as
% \begin{align}
% \label{CharCurve}
$r_c =  {R a_0(T)}/{a_0(0)}$,
% \end{align}
where $R = r_c(0) \in [0, 1)$.  Thus $M_1(r, \theta,T) = M_1(R, \theta, 0)$. In particular, $M_1(a_0, \theta,T) = M(1,\theta,0)\equiv N(\theta)$, say.  Thus the perturbation $N$ to the viscosity field at the initial {\color{black}effective} contact line location is swept forward with the contact line and maintains a long-lived influence on the flow \citep{2016-Xu-vol472}.  Recall that $M(x,y,0)\approx 1+\sigma \mathcal{G}$ for $\sigma \ll 1$ where $\mathcal{G}$ has covariance function (\ref{eq:covar}).  It follows that $N(\theta)$ has covariance 
\begin{equation}
\label{eq:covarN}
k_N(\theta,\theta')=\exp\left(-\frac{1-\cos(\theta-\theta')}{\ell^2}\right).
\end{equation}

\subsection{The inner problem}

In order to close the problem (\ref{eq:h1}) for $H_1$, we require a condition on $a_1$.  This is determined by matching to the inner problem, situated along the drop's distorted {\color{black}effective} contact line.  In the inner region, we expand using (\ref{eq:innerscale}), now allowing $a=a(\theta,T)$, recovering a modified form of the Landau--Levich equation
\begin{align}
\label{InnerLeading}
&\frac{G^3G_{XXX}}{3M} =G -1,\quad M=1+\sigma N(\theta),
\end{align}
after integrating once and considering $G \to 1$ as $ X \to \infty$.  The local viscosity field is advected with the flow from the initial condition, as explained above; we do not consider the larger-time influence of fluid that is swept into the contact line region as it advances over the undisturbed film, that may change $M$ via diffusion \citep{2016-Xu-vol472}.  The contact angle condition (\ref{CubicInner}) becomes
\begin{equation}
\Theta^3=9Ma_T (1+O(\epsilon)),
\end{equation}
which can be expanded using the condition $\epsilon \ll \sigma \ll 1$ and (\ref{eq:aexp}) to give
\begin{equation}
 \Theta^3 = 9\left(a_{0T}+\sigma (a_{1T} + a_{0T}N) +\dots\right).
\end{equation}
Matching $\Theta^3$ at order $O(1)$ and $O(\sigma)$ between the inner and outer solutions recovers $9a_{0T} = -H_{0r}^3\vert_{r=a_0}$ (yielding (\ref{eq:a0})) at leading order and
\begin{align}
\label{MatchCon}
   9  (a_{1T} + a_{0T}N) = - 3  H_{0r}^2 (a_1 H_{0rr} +H_{1r})\vert_{r = a_0},
\end{align}
at the following order. The matching condition (\ref{MatchCon}) closes the problem (\ref{eq:h1}) for the depth and contact-line perturbations $H_1(r,\theta,T)$ and $a_1(\theta,T)$.  

\subsection{Fourier series solution}

Setting $H_1 = (4V/\pi a_0^3)\overline{H_1}$, the problem for $\overline{H_1}$ in (\ref{eq:h1}) becomes
\begin{align}
\label{Poisson}
& \nabla^2 \overline{H_1} = P(T), \quad \int_0^{2\pi}  \int_0^{a_0} r \overline{H_1} \mathrm{d}r \mathrm{d}\theta = 0,
\end{align}
subject to
\begin{align}
\label{PoissonBC}
&\overline{H_1}_r - \frac{a_1}{a_0} = -\frac{1}{3}\left(N + \frac{a_{1T}}{a_{0T}}\right), \quad \overline{H_1} = a_1, \quad \text{on} \quad r = a_0.
\end{align}
Denoting $\tilde H_1 = \overline{H_1} - \tfrac{1}{4}Pr^2$, we recover a Laplace problem with a Dirichlet boundary condition
\begin{align}
&\nabla^2 \tilde H_1 = 0, \quad \tilde H_1 = a_1(\theta,T) - \tfrac{1}{4} Pa_0^2 \equiv f(\theta,T) \quad \text{on} \quad r = a_0.
\end{align}
This has the series solution
%whose solution is 
%\begin{align}
%\label{KernelSol}
%H_2(T, r, \theta) =\frac{1}{2\pi} \int_0^{2\pi}\frac{a_0^2 - r^2}{a_0^2 - 2a_0 r \cos(\theta -\phi) +r^2}f(T, \phi)\mathrm{d}\phi
%\end{align}
%or is written in a form of series solution
\begin{align}
\label{SeriesSol}
\tilde H_1(r, \theta,T) =\frac{1}{2\pi} \int_0^{2\pi}f(\phi,T)\, \mathrm{d}\phi + \frac{1}{\pi}\sum_{n=1}^{\infty}\left(\frac{r}{a_0}\right)^n \int_0^{2\pi}\cos \left[n(\theta -\phi)\right] f(\phi,T) \, \mathrm{d}\phi.
\end{align}
Substituting (\ref{SeriesSol}) into the volume constraint (\ref{Poisson}b) gives
\begin{align}
\int_0^{2\pi}  \int_0^{a_0} r \left(\frac{Pr^2}{4}+\frac{1}{2\pi} \int_0^{2\pi}f(\phi,T)\mathrm{d}\phi + \frac{1}{\pi}\sum_{n=1}^{\infty}\left(\frac{r}{a_0}\right)^n \int_0^{2\pi}\cos \left[n(\theta -\phi)\right] f(\phi,T) \mathrm{d}\phi\right) \mathrm{d}r \mathrm{d}\theta = 0,
\end{align}
which simplifies to
\begin{align}
&P(T) = \frac{4}{\pi a_0(T)^2} \int_0^{2\pi}a_1(\phi,T)\, \mathrm{d}\phi.
\end{align}
The boundary condition (\ref{PoissonBC}a) gives
\begin{align}
\label{IntegroDiff}
a_{1T} &= a_{0T}\left(3\left(\frac{a_1}{a_0} - \overline{H}_{1r}\right) - N\right)  = a_{0T}\left(3\left(\frac{a_1(\theta,T)}{a_0(T)} - \tilde H_{1r}(a_0, \theta,T) - \frac{P(T) a_0(T)}{2}\right) - N(\theta)\right) \nonumber \\
&=a_{0T}\left(\frac{3}{a_0(T)}\left(a_1(\theta,T) - \frac{1}{\pi }\sum_{n=1}^{\infty}n \int_0^{2\pi}\cos \left[n(\theta -\phi)\right] a_1(\phi,T)\, \mathrm{d}\phi - \frac{2}{\pi } \int_0^{2\pi}a_1(\phi,T)\, \mathrm{d}\phi\right) - N(\theta)\right).
\end{align}
This linear problem defines the evolution of the contact-line perturbation $a_1(\theta,T)$.  Since $a_0$ is monotonic in $T$ we may use it to parametrize time.  Writing $a_1 = a_1(\theta,a_0)$, we have
\begin{align}
\label{IntegroDiff2}
\frac{\partial a_1}{\partial a_0} =\frac{3}{a_0}\left(a_1(\theta,a_0) - \frac{1}{\pi }\sum_{n=1}^{\infty}n \int_0^{2\pi}\cos \left[n(\theta -\phi)\right] a_1(\phi,a_0) \mathrm{d}\phi - \frac{2}{\pi } \int_0^{2\pi}a_1(\phi, a_0) \mathrm{d}\phi\right) - N(\theta).
\end{align}
The special case when $N$ is independent of $\theta$, so that $a_1 = a_1(a_0)$, has the closed-form solution
\begin{align}
\label{ThetaFree}
& a_1= \frac{N(1-a_0^{10})}{10a_0^9} = -\frac{64 N V^3 T}{3^{1/5} \pi ^{3/10} \left(640 V^3 T+9 \pi ^3\right)^{9/10}},
\end{align}
where we use initial condition $a_1|_{a_0 = 1} = 0$.  Uniformly elevated viscosity $(\sigma N > 0)$ will therefore slow spreading ($\sigma a_1<0$), as expected.

Given $N(\theta)$, we can determine $a_1(\theta,a_0)$ from (\ref{IntegroDiff2}) by expressing $a_1(\theta,a_0)$ as a Fourier series
\begin{align}
\label{FourierExpansion}
a_1 = b_0(a_0) + \sum_{k = 1}^{\infty} (b_k(a_0) \cos(k \theta) + c_k(a_0) \sin(k \theta)).
\end{align}
Substituting (\ref{FourierExpansion}) into (\ref{IntegroDiff2}), we have
\begin{align}
&\frac{\mathrm{d} b_0}{\mathrm{d} a_0} + \sum_{k = 1}^{\infty} \left(\frac{\mathrm{d} b_k}{\mathrm{d} a_0} \cos(k \theta) + \frac{\mathrm{d} c_k}{\mathrm{d} a_0} \sin(k \theta)\right) = \frac{3}{a_0}\Bigg(b_0 + \sum_{k = 1}^{\infty} (b_k \cos(k \theta) + c_k \sin(k \theta)) \nonumber \\
&\qquad - \frac{1}{\pi }\sum_{n=1}^{\infty}n \int_0^{2\pi}\left(\cos (n \theta) \cos (n \phi) + \sin (n \theta) \sin (n \phi)\right) \left(b_0 + \sum_{k = 1}^{\infty} (b_k \cos(k \phi) + c_k \sin(k \phi))\right) \mathrm{d}\phi \nonumber \\
&\qquad\qquad - \frac{2}{\pi } \int_0^{2\pi} \left(b_0 + \sum_{k = 1}^{\infty} (b_k \cos(k \phi) + c_k \sin(k \phi))\right) \mathrm{d}\phi\Bigg) - N(\theta),\nonumber \\
&= \frac{3}{a_0}\Bigg(-3 b_0 + \sum_{k = 1}^{\infty}(1-k) (b_k \cos(k \theta) + c_k \sin(k \theta)) \Bigg) - N(\theta),
\end{align}
which gives
\begin{subequations}
\begin{align}
& \frac{\mathrm{d} b_0}{\mathrm{d} a_0} = - \frac{9b_0}{a_0} - \frac{1}{2\pi}\int_0^{2\pi}N(\theta) \mathrm{d} \theta, \\
&\frac{\mathrm{d} b_k}{\mathrm{d} a_0} = \frac{3(1-k)b_k}{a_0} - \frac{1}{\pi}\int_0^{2\pi}N(\theta)\cos(k \theta) \mathrm{d} \theta, \\
&\frac{\mathrm{d} c_k}{\mathrm{d} a_0} = \frac{3(1-k)c_k}{a_0} - \frac{1}{\pi}\int_0^{2\pi}N(\theta)\sin(k \theta) \mathrm{d} \theta.
\end{align}
\end{subequations}
With initial conditions $b_0(1) = b_k(1) = c_k(1) = 0$, $k = 1,\dots$, this gives
\begin{subequations}\label{FourierCoeff}
\begin{align}
\label{LeadingPerturbation}
&b_0 = \frac{(1-a_0^{10})}{10a_0^9}\frac{1}{2\pi}\int_0^{2\pi}N(\theta) \mathrm{d} \theta, \\
\label{FirstPerturbationCos}
&b_k = \frac{a_0 \left(a_0^{2-3 k}-1\right)}{3 k-2} \frac{1}{\pi}\int_0^{2\pi}N(\theta)\cos(k \theta) \mathrm{d} \theta, \\
\label{FirstPerturbationSin}
&c_k = \frac{a_0 \left(a_0^{2-3 k}-1\right)}{3 k-2} \frac{1}{\pi}\int_0^{2\pi}N(\theta)\sin(k \theta) \mathrm{d} \theta.
\end{align}
\end{subequations}
The contact line shape is given by inserting the Fourier coefficients (\ref{FourierCoeff}) into (\ref{FourierExpansion}).

\subsection{Drop statistics}
\label{sec:4}

The drop area satisfies
\begin{align}
A = \int_{\mathcal{A}} \mathrm{d}A = \int_0^{2\pi} \int_0^{a(\theta,t)}  r  \mathrm{d}r \mathrm{d}\theta = \int_0^{2\pi} \frac{a^2}{2} \mathrm{d}\theta = \pi a_0^2 + \sigma a_0 \int_0^{2\pi} a_1 \mathrm{d} \theta + \cdots  \approx \pi a_0^2 +2 \pi \sigma a_0 b_0 + \cdots.
\label{eq:areaX}
\end{align}
{\color{black}This prediction of drop area is illustrated in Figure~\ref{single_simulation_evolution}(a, dotted and dashed lines); the $b_0$ term provides a significant correction in this strongly disordered example.}  The displacement of the centre of mass from the origin $\bm{D}= D_x\bm{i} + D_y \bm{j}$ is
\begin{align}
\bm{D} &= \frac{1}{A}\int_{\mathcal{A}} \bm{r} \mathrm{d}A = \frac{1}{A}\int_{\mathcal{A}} (x \bm{i} + y \bm{j})\mathrm{d}A \nonumber \\
&= \frac{1}{A}\left(\bm{i}\left(\int_0^{2\pi} \int_0^{a(\theta,t)} r^2 \cos \theta  \mathrm{d}r \mathrm{d}\theta\right) + \bm{j}\left(\int_0^{2\pi} \int_0^{a(\theta,t)} r^2 \sin \theta  \mathrm{d}r \mathrm{d}\theta\right)\right) \nonumber \\
&= \frac{1}{A}\left(\bm{i}\left(\int_0^{2\pi}  \frac{a^3}{3} \cos \theta  \mathrm{d}\theta\right) + \bm{j}\left(\int_0^{2\pi}  \frac{a^3}{3} \sin \theta  \mathrm{d}\theta\right)\right) \nonumber \\
&= \frac{1}{A}\left(\bm{i}\left(\int_0^{2\pi}  (\sigma a_0^2a_1+\cdots) \cos \theta  \mathrm{d}\theta\right) + \bm{j}\left(\int_0^{2\pi}  (\sigma a_0^2a_1+\cdots) \sin \theta  \mathrm{d}\theta\right)\right) \nonumber \\
&= \frac{\sigma}{\pi}\left(\bm{i}\left(\int_0^{2\pi}  a_1 \cos \theta  \mathrm{d}\theta\right) + \bm{j}\left(\int_0^{2\pi}  a_1 \sin \theta  \mathrm{d}\theta\right)\right)+ \cdots \approx \sigma\left(b_1\bm{i} + c_1\bm{j}\right).
\end{align}
The magnitude of the drop displacement is $D^2 = \sigma^2 (b_1^2 + c_1^2)$.

Since $\mathbb{E}(b_0)=\mathbb{E}(b_1)=\mathbb{E}(c_1)=0$, over multiple realisations there is no systematic change to the mean drop area nor any net drift in either the $x$ or $y$ directions.  We can however test this prediction by considering the path taken by a drop in a given realisation.  According to the present approximation, 
\begin{equation}
\label{eq:driftdir}
D_x \bm{i}+ D_y\bm{j}=\frac{1-a_0}{\pi}\int_0^{2\pi}N(\theta) (\bm{i} \cos \theta + \bm{j} \sin\theta) \,\mathrm{d}\theta. 
\end{equation}
The direction of drift is therefore set by the first moment of the viscosity field around the initial location of the {\color{black}drop's effective} contact line.  The dashed line in the inset of Figure~\ref{single_simulation_evolution}(b) uses (\ref{eq:driftdir}) to provide a good prediction of the simulated direction of drift.  The drift can be expected to deviate from this prediction at larger times, as the viscosity field in the inner region is increasingly influenced by material accumulated in the contact-line region as the drop spreads over the film.

The variances of $A$ and $D$ in this approximation are
\begin{align}
\label{eq:varad}
\sigma_A^2\equiv \mathrm{Var}(A) & = (2 \pi \sigma a_0)^2 \mathrm{Var}\left( b_0\right),  \quad \mathrm{Var}(D) = \sigma^2\mathrm{Var}\left(\sqrt{b_1^2 + c_1^2}\right) .
\end{align}
In the limit $\ell \gg 1$, $N$ is independent of $\theta$ and $N \sim \mathcal{N}(0,1)$. In this case, $a_1$ is given by (\ref{ThetaFree}) and
\begin{align}
\mathrm{Var}(A) &= \left(\frac{\pi \sigma (1-a_0^{10})}{5a_0^8} \right)^2, \quad \mathrm{Var}(D) = 0.
\end{align}
The disorder in the initial viscosity field leads to variability in drop {\color{black}area} but no lateral drift.

In order to evaluate $\mathrm{Var}(b_0)$ in (\ref{eq:varad}) when $N$ is nonuniform, we define $\Delta \theta = 2\pi/K$ where $K$ is a large positive integer and approximate the integral in (\ref{LeadingPerturbation}) by a Riemann sum
\begin{align}
&\int_0^{2\pi}N_0(\theta) \,\mathrm{d} \theta = \Delta \theta \sum_{j=1}^{K} N(\theta_j),
\end{align}
where $\theta_j = (j-1) \Delta \theta$, $j = 1, \cdots, K$. Thus
\begin{align}
\mathrm{Var}(b_0) &= \left(\frac{(1-a_0^{10})}{10a_0^9}\frac{1}{2\pi}\right)^2 (\Delta \theta)^2 \sum_{j=1}^K \sum_{i=1}^K \mathrm{Cov}(N(\theta_i), N(\theta_j)) \nonumber \\
% &=\left(\frac{(1-a_0^{10})}{10a_0^9}\frac{1}{2\pi}\right)^2 (\Delta \theta)^2 \sum_{j=1}^K \sum_{i=1}^K \mathrm{Cov}(M_1(1,\theta_i,0), M_1(1,\theta_j,0)).
% \end{align}
% \begin{align}
 &=\left(\frac{(1-a_0^{10})}{10a_0^9}\frac{1}{2\pi}\right)^2 (\Delta \theta)^2 \sum_{j=1}^K \sum_{i=1}^K \exp\left(-\frac{1 - \cos(\theta_i -\theta_j)}{\ell^2}\right),
\end{align}
using (\ref{eq:covarN}).
% Now $ M|_{t=0} = \exp(\sigma\mathcal{G}(X,\omega))$, where $\mathcal{G}(X,\omega)$ is a Gaussian random field with zero mean and stationary covariance 
% \begin{align}
% k_\mathcal{G}(X, X') = \exp\left(-\frac{1}{2}\left(\frac{X-X'}{\ell}\right)^2\right).
% \end{align}
% Here $\sigma^2$ and $\ell$ are the variance and the correlation length of the Gaussian random field, respectively. Thus, in the weak disorder limit, we have $\sigma M_1 = \mathcal{G}(X,\omega)$ and therefore
As $K \to \infty$ we convert Riemann sums back to integrals, giving
\begin{align}
\mathrm{Var}(b_0) &=\left(\frac{(1-a_0^{10})}{10a_0^9}\frac{1}{2\pi}\right)^2 \int_0^{2\pi}\int_0^{2\pi} \exp\left(-\frac{1 - \cos(\theta -\theta')}{\ell^2}\right)\mathrm{d}\theta\mathrm{d}\theta', \nonumber \\
& = \left(\frac{(1-a_0^{10})}{10a_0^9}\right)^2  I_0(1/\ell^2) \exp(-1/\ell^2)
\end{align}
where $I_0(\cdot)$ is the modified Bessel function of the first kind of zeroth order.  Thus
\begin{align}\label{VarA}
\mathrm{Var}(A) & =  \left(\frac{\pi \sigma(1-a_0^{10})}{5a_0^8}\right)^2  I_0(1/\ell^2) \exp(-1/\ell^2).
\end{align}
Similarly, from (\ref{FirstPerturbationCos}, \ref{FirstPerturbationSin}), 
\begin{subequations}\label{VarDxDy}
\begin{align}
\label{VarDx}
\mathrm{Var}(D_x) & = \sigma^2\left(\frac{1-a_0}{\pi}\right)^2 \int_0^{2\pi}\int_0^{2\pi}\cos(\theta)\cos(\theta') \exp\left(-\frac{1 - \cos(\theta -\theta')}{\ell^2}\right)\mathrm{d}\theta\mathrm{d}\theta' \nonumber \\
&=2\sigma^2(1-a_0)^2I_1(1/\ell^2) \exp(-1/\ell^2), \\
\label{VarDy}
\mathrm{Var}(D_y) & = \sigma^2 \left(\frac{1-a_0}{\pi}\right)^2 \int_0^{2\pi}\int_0^{2\pi}\sin(\theta)\sin(\theta') \exp\left(-\frac{1 - \cos(\theta -\theta')}{\ell^2}\right)\mathrm{d}\theta\mathrm{d}\theta' \nonumber \\
&=2\sigma^2(1-a_0)^2I_1(1/\ell^2) \exp(-1/\ell^2),
\end{align}
\end{subequations}
where $I_1(\cdot)$ is the modified Bessel function of the first kind of first order. Furthermore, 
\begin{align}\label{MeanR2}
\mathbb{E}(D^2) = \mathbb{E}(D_x^2) + \mathbb{E}(D_y^2) = \mathrm{Var}(D_x) +  \mathrm{Var}(D_y) = 4\sigma^2(1-a_0)^2I_1(1/\ell^2) \exp(-1/\ell^2)
\end{align}
because $\mathbb{E}(D_x)=\mathbb{E}(D_y)=0$. 

\subsection{Comparison of variance predictions}
\label{sec:comp}

We now evaluate these asymptotic predictions against simulations and emulation estimates.  First, it is important to emphasise that, because of neglect of $O(\epsilon)$ corrections in the leading-order drop radius, the prediction (\ref{eq:a0}, \ref{LOCL}) lacks quantitative accuracy for moderately small values of $\eta$ {\color{black}(see Figure~\ref{fig:axisym}), although the dot-dashed line in Figure~\ref{single_simulation_evolution}(a) shows that an \textit{ad hoc} modification of the prediction using (\ref{eq:a0refined}) to determine the base-state radius instead of (\ref{eq:a0}) can be beneficial.}  Even so, the {\color{black}leading-order} predictions (\ref{VarA}, \ref{VarDxDy}) are respectable (Figure~\ref{errorbar}a-c), capturing the monotonic rise in $\sigma_A$ with respect to $\ell$ and the maxima in $\sigma_{D_x}$ and $\sigma_{D_y}$.  This result suggests that the local maximum in the emulation prediction for $\sigma_A$ is an artefact; indeed it is not evident when the training data spans a wider range of parameter space (Figure~\ref{fig:emuls2D}a).  The predicted mean-square drift (\ref{MeanR2}) works well (Figure~\ref{errorbar}d); the slight overestimate for small $\ell$ may be due to the neglect of diffusion, which suppresses gradients of the viscosity field in the contact line region \citep{2016-Xu-vol472}.

The asymptotic predictions necessarily imply that $\sigma_A$ and $\sigma_{D_y}$ are proportional to $\sigma$ when $\sigma \ll 1$, whereas emulation predictions account more accurately for the effects of strong disorder. Comparison between the two sets of predictions over $(\ell,\sigma)$-space (Figure~\ref{fig:emuls2D}) show that the asymptotic predictions strongly underestimate the magnitude of {\color{black}the} variance in drop area as $\sigma$ increases (Figure~\ref{fig:emuls2D}a,c), but overestimate the magnitude of variance in lateral drift (Figure~\ref{fig:emuls2D}b,d), although the qualitative dependence on correlation length is captured reasonably well.  

\section{Discussion}

In this study we have presented two complementary approaches for uncertainty quantification in a thin-film flow problem governed by coupled nonlinear evolution equations.  The multiscale solution structure (with pressure gradients predominantly confined to thin boundary layers) is a computational challenge, but also offers a route to analysis.  Accordingly, we used GP emulation to extract variance predictions from a modest number of simulations (without the much larger overhead of direct MC simulation) and used a low-order asymptotic (surrogate) model to derive explicit variance predictions in the weak disorder limit, {\color{black}notably (\ref{VarA}) for the drop area and (\ref{MeanR2}) for the mean square drop displacement}.

The emulation procedure {\color{black}described in Appendix~\ref{sec:gp}} is non-standard because of the requirement to capture the parameter dependence of the disorder emerging from the simulations.  This necessitated implementation of an iterative procedure due to \cite{2007-Noise}, which we deployed in our package GP\_emu\_UQSA\footnote{https://github.com/samcoveney/maGPy}.  The key advantage of this approach is that the emulator learns about the variability of outputs from simulations that randomly sample a desired region of parameter space, rather than building expensive pointwise variance estimates.

Our simulations reveal a competition between capillary and viscous effects in determining the evolution of the drop.  In the bulk of the drop, the smoothing effect of surface tension maintains the integrity of the drop and averages out the fluctuations around the contact line (Figure~\ref{single_simulation}).  This competition is embodied in the linear free-boundary problem (\ref{eq:h1}, \ref{MatchCon}) that we derived in the weak-disorder approximation, from which we derived a nonlocal evolution equation for contact-line perturbations (\ref{IntegroDiff}), forced by the non-uniform viscosity field $N(\theta)$ around the contact-line's initial location.  Restricting attention to key geometric measures of drop dynamics reduced the problem further to {\color{black}the} evaluation of a few Fourier components, defined in terms of moments of $N(\theta)$.  Prescribing the covariance structure of the initial viscosity field, we could then evaluate the statistics of these moments directly to recover variance predictions in terms of the correlation length of the initial viscosity field.  Despite well-understood limitations of the asymptotic approximation (due to slow convergence of a logarithmic expansion, {\color{black}see Appendix~\ref{sec:hior}}), these predictions capture the primary stochastic features of the simulations.  

 {\color{black}This study was motivated by considering the interaction between an inhaled aerosol droplet and the liquid lining of a lung airway.  Currently there are very limited observational data describing the spreading of aerosol droplets on realistic mucus films and little consensus in the literature on rheological parameters \citep{levy2014}; indeed, the restricted set of physical processes considered here may well be superseded by many effects that we have not considered.  Airway film thicknesses have been reported to range between 0.2 and 50 $\mu$m \citep{sims1997}, with a depth of 5$\mu$m being typical in a main bronchus.  The mucus may form discontinuous rafts, in addition to reports of mucin clouds over distances of 10$\mu$m and sheets and plumes over distances of hundreds of microns  \citep{kesimer2013, ostedgaard2017, widdicombe2015}, placing a correlation length $\ell$ potentially well in excess of the film thickness and in the range of a macroscopic droplet diameter.  Our study suggests that future models of pulmonary drug delivery (via liquid plugs or aerosols), building on studies such as those of \cite{filoche2015}, \cite{kim2015}, \cite{khanal2015} and \cite{sharma2015}, may need to account for factors such as lateral drift of spreading droplets driven by mucus heterogeneity, as described for example by (\ref{eq:driftdir}).  More generally, studies of transport problems in `dirty' environments characteristic of many biological systems will require a framework, as developed here, combining judicious model reduction and emulation techniques in order to predict the distribution of outcomes.}

\section*{Acknowledgements}

This study was supported by EPSRC grant no.~EP/K037145/1.

\noindent
The GP\_emu\_UQSA package may be downloaded from \texttt{https://github.com/samcoveney/maGPy}.

\begin{appendix}

\section{Simulations}
\label{app:sim}

We perform numerical simulations in COMSOL Multiphysics, rewriting (\ref{2d_vector_form}) {\color{black}as a system of second-order equations as}
\begin{align}\label{COMSOL}
d_a \bm{v}_t + \nabla \cdot \boldsymbol{\Gamma} = \mathbf{f},
\end{align}
where 
\begin{align}\label{COMSOL2}
d_a = \begin{pmatrix}
1&0&0 \\
0&0&0 \\
0&0&1
\end{pmatrix}, \quad
\bm{v} = \begin{pmatrix}
H \\
P \\
Q
\end{pmatrix}, \quad
\boldsymbol{\Gamma} = \begin{pmatrix}
({H^4}/{3Q}) \nabla P \\
\nabla H \\
\frac{1}{3}H^3 \nabla P - {Pe}^{-1}H \nabla (Q/H)
\end{pmatrix}, \quad
\mathbf{f} = \begin{pmatrix}
0 \\
P \\
0
\end{pmatrix}
\end{align}
{\color{black}where $Q\equiv HM$}.  The computational domain is a square $[-L, L] \times [-L, L]$ around which a no-flux boundary condition $\bm{n}\cdot{\boldsymbol{\Gamma}} =0 $ is applied, where $\bm{n}$ is the outward-facing normal.  Simulations are terminated before the boundary has any direct influence.  {\color{black}We used COMSOL's built-in mesh generator to create an unstructured triangular mesh on the computational domain and its default second-order elements to discretize (\ref{COMSOL}, \ref{COMSOL2}).}  Convergence tests were undertaken to ensure that reported results were grid-independent. {\color{black}The 2D scheme was also tested against a 1D finite-difference solver that used the method of lines and the Matlab time-stepping routine \texttt{ode23t} in an axisymmetric problem (see the solid and dotted lines in Figure~\ref{fig:axisym}; both methods quickly regularise the initial pressure jump in (\ref{eq:ic}).)}.  We used the routines $\mathtt{g05zr}$ and $\mathtt{g05zs}$ in the NAG toolbox for MATLAB to generate realizations of the stationary Gaussian random field $\mathcal{G}$ {\color{black}with covariance (\ref{eq:covar})} by circulant embedding \citep{Lord2014}.  Integrals in (\ref{qoi_definition}) were evaluated by defining $\mathcal{A} = \{(x,y):H(t,x,y)>1.01\eta\}$.  We collected statistics from simulations initiated from multiple realisations of the initial viscosity field in order to characterise the distribution of outcomes.

\section{Emulation}
\label{sec:gp}

A GP emulator of the simulation outputs (implemented in the Python package GP\_emu\_UQSA\footnote{https://github.com/samcoveney/maGPy}) was constructed following \cite{2007-Noise}.  The simulator is modelled as a noisy regression problem with inputs $\mathbf{x}_i$ (representing model parameters in realisation $i$) leading to a noisy output $y_i$, of the form
\begin{align} \label{eq:noisy_outputs}
	y_i = f(\mathbf{x}_i) + \delta_i,\quad \delta_i\sim \mathcal{N}(0,r(\mathbf{x}_i)),\quad i=1,\dots,N.
\end{align}
Here $f$ represents the nonlinear model and $r$ the heteroscedastic noise in the model (\hbox{i.e.} the parameter-dependent variability that is propagated forwards from the initial conditions to the final state).  For datasets of each quantity of interest $\{y_i\}$ (\hbox{i.e.} multiple predictions of $A(t_f)$ or $D_y(t_f)$), the functions $f(\mathbf{x})$ and $r(\mathbf{x})$ must be learned from training data.  GP priors are placed on $f$ and $r$, using exponential covariance functions with hyperparameters (variances and correlation lengths) that are also learned from training data.  Once determined, the model provides a posterior distribution for outputs $y^*_j$ at parameters $\mathbf{x}_j^*$ as a multivariate normal distribution.  The latent variables ($r(\mathbf{x}_i)$) are determined using a most-likely approach \citep{2007-Noise}.  The hyperparameters are determined using an expectation-maximization algorithm, in which a GP is first trained on the simulation data $\{\mathbf{x}_i, y_i\}$ using a prediction for the mean of the variance at each point $\{ \mathbb{E}[r_i] \}$ (initially set to zero as unknown). This `mean GP' is used to estimate the actual pointwise variance $\{ r_i \}$ at each $\mathbf{x}_i$, and a different `noise GP' is then trained on the $\{\mathbf{x}_i, z_i\}$, where $z_i = \log(r_i)$, in order to obtain a smooth estimate of the mean of the noise $\{ \mathbb{E}[r_i] \}$, which allows the `mean GP' to be retrained. This iterative procedure is repeated until convergence.  In this problem, because the variation of the mean function is much smaller than the variance arising from the random initial conditions (unlike most machine-learning problems, where the variance of the signal typically exceeds that of the noise), learning the exact form of the mean function was not possible. To ensure that the `mean GP' {\color{black}fitted} the data well, a random sample from a noiseless GP was added to the data, preserving the noise levels; adding different samples preserved the results of fitting the noise.

\section{Axisymmetric spreading in the uniform viscosity limit}
\label{sec:hior}

{\color{black}

\begin{figure}
\begin{center}
\begin{overpic}{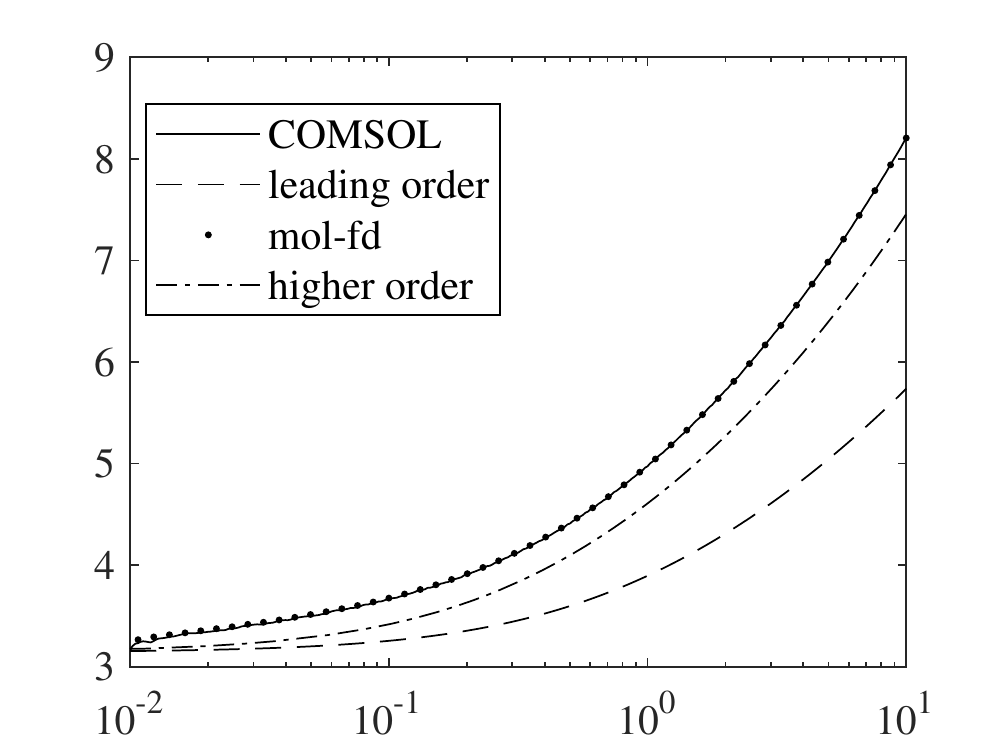}
\put(50,-5) {$t$}
\put(-2,40) {$A$}
\end{overpic}
\caption{Axisymmetric drop spreading with uniform viscosity ($M=1$), with $V=\pi/2$ and $\eta=0.01$.  The solid and dotted lines (which are almost indistinguishable) compare respectively the 2D COMSOL solver and a 1D finite difference code.  The dashed and dot-dash lines compare respectively the leading-order approximation (\ref{eq:a0}) and the refined approximation (\ref{eq:a0refined}).}
\label{fig:axisym}
\end{center}
\end{figure}

We revisit here the classical problem (\ref{eq:hmu}) with uniform viscosity ($M=1$), for which an initially axisymmetric drop remains axisymmetric, to illustrate how pressure gradients across the bulk of the drop influence spreading rates.  Setting $H = H(r, T)$, where $r$ is the radial coordinate, we match solutions in the inner and outer regions as before
%\begin{align}
%\epsilon H_T + \frac{1}{3 r} \left(r H^3 \left(\frac{1}{r}(rH_r)_r
%%\frac{\partial^3 H}{\partial r^3} + \frac{1}{r}\frac{\partial^2 H}{\partial r^2} - \frac{1}{r^2}\frac{\partial H}{\partial r}
%\right)_r\right)_r = 0
%\end{align}
in the limit $\epsilon\rightarrow 0$.  In the outer region we take $H\approx H_0+\epsilon H_{(1)}+\dots$, $\mathbf{U}\approx \mathbf{U}_0+\epsilon \mathbf{U}_{(1)}+\dots$, where $H_0$ and $\mathbf{U}_0$ are given by (\ref{eq:axisol}).  The correction to the film thickness satisfies
\begin{equation}
\frac{a_T}{a} r = \frac{H_0^2}{3} \left( \frac{1}{r}\left( rH_{(1)r}\right)_r \right)_r
\end{equation}
where we apply $H_{(1)}\rightarrow 0$ as $r\rightarrow a$, $H_{(1)r}\rightarrow 0$ as $r\rightarrow 0$ and $\int_0^a rH_{(1)}\,\mathrm{d}r=0$.  Integrating, one finds that
\begin{equation}
H_{(1)r}=\frac{3\pi^2 a^7 a_T}{16 V^2} \left[\zeta r - \frac{\log\left(1-(r^2/a^2)\right)}{r} \right]
\end{equation}
for some constant $\zeta$.   Integrating the volume constraint by parts reveals that $\zeta=-2/a^2$.  Near the contact line, it follows that
\begin{equation}
H_r^3\approx -\left(\frac{4V}{\pi a^3} \right)^3 -9 \epsilon a_T \left(2+\log\left(1-(r/a)\right) + \log 2\right) + \dots
\end{equation}
Matching with (\ref{CubicInner}) and writing $a_t=\epsilon a_T$ leads to
\begin{align}
\label{eq:a0refined}
 \left(\frac{4V}{a^3 \pi}\right)^3 =   9  a_t  \left( \ln (1/\eta) + \ln (a/2)  + \tfrac{1}{3} \ln (3a_t) - 1  + 3^{2/3} \alpha \right).
\end{align}
Retaining the leading-order term as $\eta\rightarrow 0$ on the right-hand-side of (\ref{eq:a0refined}) recovers (\ref{eq:a0}).  The higher-order corrections are analogous to those derived (for slip) by for example \cite{hocking1983} and other authors.

The corrections in (\ref{eq:a0refined}) provide an important boost to the accuracy of this approximation.  Figure~\ref{fig:axisym} illustrates the point, showing an axisymmetric drop-spreading simulation for $\eta=0.01$ using (\ref{COMSOL}) and comparing it to (\ref{eq:a0}) and the refined prediction (\ref{eq:a0refined}).  The planar analogue of (\ref{eq:a0refined}) is given in \cite{2016-Xu-vol472}; figure 2 of that paper illustrates much closer agreement between asymptotics and numerics once $\eta$ is reduced by a factor of 10 to 0.001.  Our numerical scheme (for nonaxisymmetric spreading) could not resolve such a thin precursor film with the resources available to us.   Significantly, the higher-order corrections in (\ref{eq:a0refined}) are not essential in order to capture the dominant effects of non-uniform viscosity such as the direction of drift (Figure 2b) and the dependence of variance in drop area and drift on correlation length (Figure 3).

}

\end{appendix}

\bibliographystyle{apalike}
\addcontentsline{toc}{section}{\refname} % add \refname (that is References for article class) to toc
\bibliography{References} 
\end{document}